\documentclass[12pt]{iopart}
\usepackage{iopams}
\usepackage{theorem}
\usepackage{algorithm}
\usepackage{graphicx}
\usepackage[usenames]{color}
\usepackage{multirow}

\newcommand{\eqref}[1]{\eref{#1}}

  {\end{list}}

\makeatletter

  \gdef\listctr{list\romannumeral\the\@listdepth}\expandafter
  
\makeatother

\newtheorem{theorem}{Theorem}[section]

\newenvironment{AlgorithmSteps}[1][1]{%
  \begin{list}{\csname label\listctr\endcsname}{%
      \usecounter{\listctr}
      
      \settowidth{\labelwidth}{\textsc{Step\ #1.}}%
      \setlength{\leftmargin}{\labelwidth}\addtolength{\leftmargin}{\labelsep}}}%
  {\end{list}}

\newcommand{\ve}[1]{\boldsymbol{#1}}

\newcommand{\R}{\mathbb{R}}

\newcommand{\for}{{\tilde f}}
\newcommand{\hor}{{\tilde h}}
\newcommand{\hori}{\ve {\tilde h}_{\ve i}}
\def\xk{\ve x^{(k)}}

\begin{document}

\title[Blind deconvolution in post-adaptive-optics astronomical imaging]
{A convergent blind deconvolution method for 
post-adaptive-optics astronomical imaging}

\author{ M Prato$^1$, A La Camera$^2$,  S Bonettini$^3$ and M Bertero$^2$}
\address{$^1$ Dipartimento di Scienze Fisiche, Informatiche e Matematiche, 
Universit\`a di Modena e Reggio Emilia, Via Campi 213/b, 41125 Modena, Italy}
\address{$^2$ Dipartimento di Informatica, Bioingegneria, Robotica e 
Ingegneria dei Sistemi, Via Dodecaneso 35, 16145 Genova, Italy}
\address{$^3$ Dipartimento di Matematica e Informatica, 
Universit\`a di Ferrara, Via Saragat 1, 44122 Ferrara, Italy}
\eads{\mailto{marco.prato@unimore.it}, \mailto{andrea.lacamera@unige.it}, 
\mailto{silvia.bonettini@unife.it}, \mailto{bertero@disi.unige.it}}

\begin{abstract}
In this paper we propose a blind deconvolution method which
applies to data perturbed by Poisson noise. The objective 
function is a generalized Kullback-Leibler (KL) divergence, depending on 
both the unknown object and unknown point spread function (PSF), without
the addition of regularization terms; constrained minimization, with
suitable convex constraints on both unknowns, is considered. The problem 
is non-convex and we propose to solve it by means of an inexact alternating
minimization method, whose global convergence to stationary points of
the objective function has been recently proved in a general setting.
The method is iterative and each iteration, also called
outer iteration, consists of alternating an update of the object and the 
PSF by means of fixed numbers of iterations, also called inner 
iterations, of the scaled gradient projection (SGP) method.
Therefore the method is similar to other proposed methods based on
the Richardson-Lucy (RL) algorithm, with SGP replacing RL. The use of 
SGP has two advantages: first, it allows to prove global convergence 
of the blind method; secondly, it allows the introduction of different 
constraints on the object and the PSF. The specific constraint  
on the PSF, besides non-negativity and normalization, is an upper bound 
derived from the so-called Strehl ratio (SR), which is the ratio between 
the peak value of an aberrated versus a perfect wavefront. Therefore a 
typical application, but not the unique one, is to the imaging of
modern telescopes equipped with adaptive optics systems for partial
correction of the aberrations due to atmospheric turbulence.
In the paper we describe in detail the algorithm and we recall the
results leading to its convergence. Moreover we illustrate its
effectiveness by means of numerical experiments whose results 
indicate that the method, pushed to convergence, is very promising in 
the reconstruction of non-dense stellar clusters. The case of more complex 
astronomical targets is also considered, but in this case regularization 
by early stopping of the outer iterations is required. However the proposed
method, based on SGP, allows the generalization to the case of 
differentiable regularization terms added to the KL divergence, even if
this generalization is outside the scope of this paper.
\end{abstract}

\section{Introduction}

Blind deconvolution is the problem of image deblurring when the blur is
unknown and, in general, is investigated by assuming a space-invariant model; 
in such a case the naive problem formulation is to solve the equation 
\begin{equation*}
\ve g= \ve h* \ve f~~,
\end{equation*}
where $\ve g$ is the detected image and 
$(\ve f, \ve h)$ are respectively the unknown object and the unknown point 
spread function (PSF), while $*$ denotes convolution. It is obvious that this 
problem is extremely undetermined and that there is an infinite set of 
pairs solving the equation. Among them also the trivial solution 
$(\ve f = \ve g~,~\ve h= \ve \delta)$, where $\ve \delta$ denotes the usual
delta function. Therefore the problem must be reformulated by introducing as 
far as possible all available {\it a priori} information on both the object 
and the PSF. 
 
Blind deconvolution is the subject of a wide literature and we do not
try to give a thorough account for that. Indeed the different approaches 
concern specific classes of images and PSFs. For instance approaches
applicable to natural images may not be suitable in microscopy;
approaches developed for motion blur are not applicable to other classes of
blur, and so on. As concerns natural images we only mention a
recent paper \cite{levin} which contains a critical analysis as well as 
several relevant references. 

In this paper we focus on astronomical imaging by assuming that an adaptive 
optics (AO) system is used to compensate for atmospheric blur and that a 
parameter characteristic of this correction, the so-called Strehl ratio (SR), 
is approximately known. We recall that SR is the ratio of peak diffraction 
intensity of an aberrated versus perfect waveform. In the case of AO images 
this parameter can be estimated by the astronomers during the observation 
and provided with an error of few percent (about 4-5 \%). Since this 
information provides an upper bound on the maximum value of the PSF, it can 
be used to exclude the trivial solution mentioned above and corresponding to 
the pair $(\ve g,\ve \delta)$. 

The approach we propose applies to astronomical imaging if noise is dominated 
by photon counting and therefore the data are realizations of Poisson 
processes, even if approaches based on regularized least-square methods are 
also available (see for instance \cite{jefferies}). In the Poisson 
case several iterative methods have been already investigated, which consist
of alternating updates of the object and PSF by means of Richardson-Lucy (RL) 
iterations \cite{holmes,fish,lanteri1994,tsumuraya,desidera2006,desidera2009}, 
or accelerated RL iterations \cite{biggs}. 

In \cite{holmes} one iteration of the algorithm consists of updating both 
the object and the PSF by means of one RL iteration. This algorithm was 
investigated, in a different context, by Lee and Seung \cite{lee2001} but 
their convergence proof is incomplete, since only the monotonic decrease of 
the objective function is shown while, for a general descent method to
be convergent, strongest Armijo-like decreasing conditions have to be 
verified \cite{Nocedal}. The other approaches could be classified as 
methods of inexact alternating minimization since they use alternately a 
number of RL iterations on both the object and the PSF (remark, however, 
that the optimization problem underlying these approaches is not explicitly 
mentioned by the authors). Their convergence is not proved if RL, or the 
acceleration of RL proposed in \cite{biggs}, are the algorithms used for 
inexact optimization.

In a recent paper \cite{bonettini2011}, in the context of non-negative 
matrix factorization, convergence of inexact alternating minimization
is proved if the iterative algorithm used for the inner iterations
satisfies suitable conditions, which are satisfied by the scaled
gradient projection method (SGP) \cite{bonettini2009} proposed for 
constrained minimization of convex differentiable functions. Therefore
in this paper we utilize these results to propose a convergent
blind deconvolution approach applicable to the reconstruction of 
astronomical adaptive-optics (AO) corrected images. We remark that the
advantage of SGP is not only a fast convergence, if a suitable scaling
of the gradient is used, but also the possibility of introducing
suitable convex constraints on the solution. The practical limitation is
that the projection operator onto the convex set defined by the constraints
should be easily computable. This is the case of box and equality 
constraints and the constraints we introduce on the PSF just belong to
this class. Remark that, in the case of Gaussian noise, a similar
situation is achievable if the projected Landweber method is used
(for an application to seismology see \cite{bertero1998}) or, more
precisely, the accelerated version provided by the application of gradient
projection methods \cite{benvenuto2010}. Indeed, in this case the
conditions required in \cite{bonettini2011} for convergence are
satisfied. 

The structure of our approach is similar to that of the previously mentioned
methods based on the RL algorithm, the difference being of course that RL
is replaced by SGP with different constraints on $\ve f$ and $\ve h$: in
the case of the object we only consider non-negativity while in the case of
the PSF we consider both non-negativity and an upper bound provided by the
knowledge of the SR, as well as the normalization condition which must be
satisfied by the PSF. We point out that the relevance of the use of the
SR constraint for blind deconvolution was first pointed out by Desider\`a
\& Carbillet \cite{desidera2009} and this paper intends to use it in a
proper mathematical context. 

Thanks to \cite{bonettini2011} the convergence is 
assured if we use a fixed number of SGP iterations for updating the object 
and the PSF; the number of iterations may be different in the two cases (for 
the denomination {\it asymmetric iterative blind deconvolution}, see
\cite{biggs}). Since the problem is non-convex, the limit of
the iteration may depend on the choice of the initial step and possibly
on the numbers of internal iterations. The convergence result does not 
assure that the limit is a sensible solution of the problem, since we do not 
introduce regularization in our approach. A comment on this point is required.

In the case of deconvolution it is well known that the minimizers of the
discrepancy function for Poisson data, the generalized Kullback-Leibler 
(KL) divergence, are sparse objects, i.e. they consist of bright spots over
a black background; it is the so-called {\it night-sky} \cite{barrett} or 
{\it checker-board} \cite{natterer} effect. As a result these minimizers  
can be sensible solutions in the case, for instance, of the deconvolution
of images of not dense star clusters by a given PSF and this result is 
confirmed by a wide set of numerical experiments. On the other hand, if the 
data is the image of a star cluster and we deconvolve it using a sparse 
object with points correctly located at the positions of the stars, we may 
expect that the result is a satisfactory reconstruction of the PSF; this 
reconstruction should improve as the reconstructed image used in the 
deconvolution improves. Therefore, using a suitable strategy in the choice 
of the number of inner iterations, we can expect sensible results in the 
case of stellar objects, by pushing to convergence the outer iterations. 
This argument is supported by our numerical experiments.

The situation is different in the case of diffuse or complex objects. In
this case we can believe that the semi-convergent behaviour of RL or SGP or 
similar methods implies a similar behaviour for the outer iterations of
the blind method; and this is just what we find in our tests. An 
alternative is obviously the introduction of regularization by adding 
suitable penalty terms to the KL divergence. However there are two main 
problems: the first is the selection of a suitable regularization for the 
astronomical object to be reconstructed and of a suitable regularization 
for the PSF to be reconstructed; the second is the selection of a suitable 
rule for estimating the two regularization parameters. We do not know 
generally accepted solutions for both problems and therefore early stopping 
of the iterations is still the easiest approach to regularization. We only 
remark that the proposed method can be easily generalized to the case of 
differentiable penalties thanks to the generalization of SGP to the 
regularized case, as proposed in several papers 
\cite{zanella,stagliano,bonettini2011a}.

The paper is organized as follows. In Sect. 2 we formulate the blind 
deconvolution problem as a constrained minimization of the generalized 
KL divergence 
as follows from a maximum-likelihood approach to Poisson data deblurring 
\cite{bertero2009}. In Sect. 3 we summarize the results on the inexact 
alternating minimization problem proved in \cite{bonettini2011} as well the 
main features of the SGP method proposed in \cite{bonettini2009}. In both 
cases we also provide the algorithms used in this paper. Finally in Sect. 4 
we describe our numerical experiments with a particular focus on the case of 
astronomical objects consisting of small star clusters, represented in our 
simulations by point-wise objects. In these cases we observe a remarkable 
convergence of the reconstructed PSF to that used in image simulations. We 
also attempt an accurate analysis of the artifacts generated in the 
reconstructed images since an understanding of their structure may be 
important in the practical applications of our method. Sect. 5 is devoted to 
a few conclusions and possible extensions. 

\section{Problem setting}

Following \cite{snyder1993}, we assume that the observed image 
$\ve g$ can be modeled as the sum of two terms 
$\ve g = \ve g_{pe} + \ve r~$. The first, $\ve g_{pe}$, is 
the number of photo-electrons due to object and background emission
and is a realization of a Poisson random variable with expected value 
$\ve {\tilde g} =\ve \hor \ast \ve \for+\ve b$, where $\ve \for $ is the 
original object, $\ve \hor $ is the PSF of the acquisition system and 
$\ve b$ is the background term, while $\ve r$ represents the read-out noise 
(RON). Here and in the following we denote by bold letters $N\times N $ 
arrays whose pixels are indexed by a multi-index $\ve i=(i_1,i_2)$, 
$\ve i \in S$. For simplicity we assume that the background is constant 
and known. As concerns the RON, it is a realization of a Gaussian additive 
random variable with a known variance $\sigma^2$. According to Snyder et al.
\cite{snyder1995}, it can be approximated by a Poisson process with
mean and variance being the same as $\sigma^2$ if the constant term 
$\sigma^2$ is added to $\ve g$. If we add $\sigma^2$ also to the background
and if, with an abuse of notation, we denote again as $\ve g$ and $\ve b$ 
the modified image and background, then we can conclude that
\begin{equation*}  
\ve g \sim \mbox{Poiss}(\ve \hor \ast \ve \for + \ve b)~~.
\end{equation*}
As concerns the PSF, we assume that it is normalized to unit volume
\begin{equation}
\sum_{\ve i\in S}  \hori = 1 
\label{norm}
\end{equation}
and that its maximum value, denoted by $s$, is known
\begin{equation}
\max_{\ve i\in S} \hori = s.
\label{strehl}
\end{equation}
The upper bound $s$ can be obtained by computing the diffraction-limited 
PSF of the considered telescope and multiplying its peak value by the SR
value provided by the astronomers, as discussed in the Introduction.

The blind deconvolution problem consists in finding an approximation of both 
$\ve \for$ and $\ve \hor$, given $\ve g$, $\ve b$ and $s$. To this
purpose we consider a maximum-likelihood approach to the problem of
image deconvolution. Since the maximization of the likelihood, which 
depends on the unknown object and PSF, is equivalent to the minimization
of a generalized KL divergence, we propose to estimate these approximations 
by minimizing this function (see the comments in the Introduction
concerning regularization) while taking into account all the available 
information, i.e. the non-negativity of both the PSF and the original object 
and the constraints \eqref{norm}-\eqref{strehl}. The resulting 
optimization problem is the following
\begin{eqnarray}
\label{prob}
& &{\rm min}~~KL(\ve g,\ve h\ast \ve f+\ve b) \\ \nonumber
& &{\rm s.t.}~~~\ve f\geq 0~;~ 0\leq \ve h\leq s~,\sum_{\ve i\in S} \ve{h_i} = 1
\end{eqnarray}
where $KL$ denotes the KL divergence of $\ve h\ast \ve f + \ve b$ from $\ve g$
\begin{equation}
\label{KL_divergence}
KL(\ve g,\ve h\ast \ve f + \ve b) =
\sum_{\ve i\in S} \left\{\ve{g_i}\log\frac{\ve{g_i}}{(\ve h\ast \ve f)_{\ve i} +
\ve{b_i}}+(\ve h\ast \ve f)_{\ve i} +\ve{b_i}-\ve{g_i}\right \}~~.
\end{equation}
Problem \eqref{prob} is convex if restricted to $\ve f$ or $\ve h$ only, but 
is in general non-convex with respect to the pair $(\ve f,\ve h)$, thus 
leading to the possible presence of several local minima. Indeed, the gradient 
and Hessian of the objective function in 
\eqref{prob} are given by
\begin{equation*}
\begin{array}{l}
\nabla_f KL(\ve g,\ve h\ast \ve f + \ve b) = {\ve 1}-H^T \frac{\ve g}{\ve h\ast \ve f + \ve b} \vspace*{0.3cm}\\
\nabla_h KL(\ve g,\ve h\ast \ve f + \ve b) = F^T{\ve 1}-F^T \frac{\ve g}{\ve h\ast \ve f + \ve b} \vspace*{0.3cm}\\
\nabla^2 KL(\ve g,\ve h\ast \ve f + \ve b) =\vspace*{0.3cm}\\
\ \ \ \left(\begin{array}{cc} H^T\mbox{diag}\left(\frac{\ve g}{(\ve h\ast \ve f + \ve b)^2} \right)H &
H^T\mbox{diag}\left(\frac{\ve g}{(\ve h\ast \ve f + \ve b)^2} \right)F-K(\ve h,\ve f)\\
F^T\mbox{diag}\left(\frac{\ve g}{(\ve h\ast \ve f + \ve b)^2} \right)H-K(\ve h,\ve f)^T & F^T\mbox{diag}\left(\frac{\ve g}{(\ve h\ast \ve f + \ve b)^2} \right)F\vspace*{0.3cm}\end{array}\right)
\end{array}
\label{hessian}
\end{equation*}
where $H$ and $F$ are the Block Circulant with Circulant Blocks (BCCB) 
matrices associated to the convolution, i.e. 
$\ve h\ast \ve f = H\ve f= F\ve h$, while $K(\ve h,\ve f)$ is the Block 
Hankel with Hankel Blocks (BHHB) matrix whose last row is the vector 
$\frac {\ve g}{\ve h\ast \ve f + \ve b}$ 
(see \cite[Chapter 4]{Hansen-etal-2006} for a survey on structured matrices).
Here the ratios and the squares are computed element-wise, and ${\ve 1}$ is a 
column vector with all entries equal to 1.
Even if the diagonal blocks of the Hessian are symmetric positive 
semi-definite, the whole matrix is difficult to analyze and compute.

\section{Alternating Minimization}

Despite the complexity of the Hessian, the constraints have a simple, 
separable structure, which can be exploited by adopting an Alternating 
Minimization (AM) algorithm for the solution of the non-convex problem
\eqref{prob}-\eqref{KL_divergence}. More precisely, the AM algorithms can be 
applied to any problem of the form
\begin{eqnarray}
\label{minAM}
& &{\rm min}~~J(\ve x)\\ \nonumber
& &{\rm s.t.}~~~\ve x\in \Omega_1\times\Omega_2\times...\times\Omega_m\subseteq \R^n
\end{eqnarray}
where, for all $i=1,...,m$, $\Omega_i$ is a closed and convex subset of 
$\R^{n_i}$ with $n_1+...+n_m = n$ and
any vector in the feasible set can be partitioned into vector components as
$ \ve x = (\ve x_1,\ve x_2,...,\ve x_m) \ \ \ \ve x_i \in \Omega_i$. Clearly, 
the blind deconvolution problem \eqref{prob} is a special case of 
\eqref{minAM} with $m=2$, $\ve{x}_1 = \ve{f}$ and $\ve{x}_2 = \ve{h}$. 

The basic idea of AM is the cyclic minimization of the objective function with 
respect to one variable, updating its value for the next optimization steps: 
in particular, AM is often referred to as the 
{\it Nonlinear Gauss-Seidel (GS) method}, where the iterate
$ \ve x^{(k+1)} = (\ve x_1^{(k+1)},...,\ve x_m^{(k+1)})$  is computed such that
for $i=1,...,m$ the block of variables
$\ve x_i^{(k+1)}$ is a solution of the sub-problem
\begin{eqnarray}
\label{GS}
& &{\rm min}~~J(\ve x_1^{(k+1)},...,\ve x_{i-1}^{(k+1)},\ve y,\ve x_{i+1}^{(k)},...,\ve x^{(k)}_m)~. \\ \nonumber
& &{\rm s.t.}~~~\ve y\in\Omega_i
\end{eqnarray}
This kind of approach has been widely studied in the literature 
\cite{Bertsekas,Bertsekas-Tsitsiklis88,Grippo-Sciandrone-1999,Grippo-Sciandrone-2000,LuoTseng91,Tseng91} and we recall two important facts about it:
\begin{itemize}
\item for $m=2$ it has been proved in \cite[Corollary 2]{Grippo-Sciandrone-2000} that the limit points of the sequence $\{\xk\}$ defined in \eqref{GS} are stationary for problem \eqref{minAM} even in the non-convex case;
\item for $m\geq 3$ the convergence of the nonlinear GS method \eqref{GS} to a solution of \eqref{minAM} is not guaranteed, without additional convexity assumptions on the objective function $J$: indeed, in \cite{Powell}, Powell devises a counterexample with $m=3$ where all the limit points of the sequence generated by the nonlinear GS method are not stationary for the problem \eqref{minAM}. Some convergence results are proved for example in \cite{Bertsekas, Bertsekas-Tsitsiklis88,Grippo-Sciandrone-2000,LuoTseng91} under suitable strict convexity assumptions.
\end{itemize}
All the convergence results mentioned above, even in the case $m=2$, are 
proved when the iterates are updated by an {\it exact} solution of the partial 
minimization problem \eqref{GS}, which is often impractical or too costly to 
compute. Indeed, many practical AM algorithms, which are also referred to as 
{\it Block Coordinate Descent methods}, are obtained by applying an iterative 
minimization method to approximately solve \eqref{GS}. In this case, the 
convergence properties of the alternating scheme also depend on the features of 
the inner solver.
A detailed analysis of the Block Coordinate Descent algorithms in the 
unconstrained case is proposed in \cite{Grippo-Sciandrone-1999}, where the 
authors devise some convergence conditions not necessarily related to the 
convexity of the objective function.

In this paper, we follow the approach in \cite{bonettini2011}, where the 
partial minimization over each variable \eqref{GS} is performed {\it inexactly} 
by means of a fixed number of Scaled Gradient Projection (SGP) steps 
\cite{bonettini2009}.
\begin{equation}
\begin{array}{l}
\mbox{Choose a feasible starting point } \ve x^{(0)} \mbox{ and a positive integer } L\geq 1\\
\mbox{For }k=0,1,2,...\\
\left\lfloor  
\begin{array}{l} \mbox{For }i=1,...,m \\ 
\left\lfloor
\begin{array}{l} \mbox{Compute } \ve x_i^{(k+1)} \mbox{ by applying } n_i^{(k+1)}\leq L \mbox{ SGP iterations to \eqref{GS}}
\end{array}\right.
\end{array}\right.\\
\end{array}
\label{CBGP}
\end{equation}

A representation of the scheme \eqref{CBGP} applied to the blind deconvolution 
problem is given in Algorithm \ref{CSGP}: each main cycle consists of two 
successive deconvolution steps to update the current estimates of the object 
$\ve f^{(k)}$ and PSF $\ve h^{(k)}$. For sake of completeness, we report the 
convergence result shown in Theorem 4.2 of \cite{bonettini2011} for our 
particular case.
\begin{theorem}
Every limit point of the sequence $(\ve f^{(k)},\ve h^{(k)})$ generated by 
Algorithm \ref{CSGP} is a stationary point for problem \eqref{prob}.
\end{theorem}

As far as we know, convergence results stronger than that given in this 
Theorem for first-order methods applied to a general non-convex problem 
do not exist in the literature. The main difficulty in this kind of problems 
is to prove the existence of convergent sub-sequences. However, in our 
specific case, thanks to the Strehl constraint, the sequence of the PSFs 
generated by our approach is bounded. Moreover, as concerns object 
reconstruction one could introduce a constraint on the flux ($\ell_1$ norm) 
of the reconstructed objects since this parameter can be derived from the 
data. However, as follows from our numerical experience this constraint is 
practically assured by SGP (we recall that it is exactly assured by RL in 
the case of zero background) and therefore we do not introduce it in order 
to reduce the computational cost. It follows that also the sequence of the 
reconstructed objects is bounded. We conclude that the existence of 
convergent sub-sequences is assured. We can add that we remarked a convergent 
behaviour of all the sequences obtained in our numerical experiments.

Finally, we point out that the convergence result holds for any number of 
inner SGP iterations. The key point of such theoretical analysis is the 
sufficient decrease of the objective function, which is enforced at each 
SGP iteration by means of an Armijo backtracking loop. Since the 
objective function \eqref{KL_divergence} is not convex with respect to 
the couple $(\ve f,\ve h)$, the presence of multiple potential stationary 
points makes any limit point dependent on both the initial guess and the 
chosen inner iteration numbers on the image ($n_f$) and the PSF ($n_h$).

\begin{algorithm}[ht]
\caption{Cyclic Scaled Gradient Projection (CSGP) method}
\label{CSGP}
Choose the starting point $\ve f^{(0)}, \ve h^{(0)}$ and the inner iterations numbers $n_f,n_h\geq 1$.\\[.2cm]
{\textsc{For}} $k=0,1,2,...$ \textsc{do the following steps:}
\begin{itemize}
\item[]
\begin{AlgorithmSteps}[4]
\item[1] Compute $\ve f^{(k+1)}$ with $n_f$ SGP iterations applied to
\begin{eqnarray}
\label{probf}
& &{\rm min}~~KL(\ve g,\ve h^{(k)}\ast \ve f + \ve b) \\ \nonumber
& &{\rm s.t.}~~~\ve f\geq 0
\end{eqnarray}
starting from the point $\ve f^{(k)}$
\item[2] Compute $\ve h^{(k+1)}$ with $n_h$ SGP iterations applied to
\begin{eqnarray}
\label{probh}
& &{\rm min}~~KL(\ve g,\ve h\ast \ve f^{(k+1)}+\ve b) \\ \nonumber
& &{\rm s.t.}~~~0\leq \ve h\leq s~,\sum_{\ve i\in S} \ve{h_i} = 1
\end{eqnarray}
starting from the point $\ve h^{(k)}$.
\end{AlgorithmSteps}
\end{itemize}
\textsc{End}
\end{algorithm}

We stress the fact that the main strength of Algorithm \ref{CSGP} for blind 
deconvolution with respect to the more standard AM approach described for 
example in \cite{Chan-Wong-2000}, is that it allows an inexact solution of 
the inner sub-problems \eqref{probf}--\eqref{probh} while preserving the 
theoretical convergence properties. Since the proposed method is essentially 
based on SGP, we recall its main features in the following sub-section.

\subsection{The Scaled Gradient Projection method}

The SGP algorithm is a first--order method which applies to any optimization 
problem of the form
\begin{equation}
\min_{\ve x\in \Omega} J(\ve x)
\label{minprob}
\end{equation}
where $J(x)$ is a continuously differentiable function and $\Omega$ is a convex set. Each SGP iteration is based on the feasible descent direction defined as
\begin{equation*}
\ve d^{(k)} = P_{\Omega,D_k^{-1}}(\ve x^{(k)}-\alpha_kD_k\nabla J(\ve x^{(k)})) -\ve x^{(k)}
\end{equation*}
where $\alpha_k$ is a scalar step-size parameter, $D_k$ is a diagonal matrix with positive diagonal entries and $P_{\Omega,D_k^{-1}}(\cdot)$ is the projection onto $\Omega$ associated to the norm induced by $D_k^{-1}$, i.e.
\begin{equation}P_{\Omega,D_k^{-1}}(\ve x)= \mbox{arg}\min_{\ve y\in\Omega} (\ve x-\ve y)^TD_k^{-1}(\ve x-\ve y). \label{proj}\end{equation}
The new point is computed along the direction $\ve d^{(k)}$ as follows
\begin{equation*}
\ve x^{(k+1)} = \ve x^{(k)} +\lambda_k\ve d^{(k)}
\end{equation*}
where $\lambda_k$ is a step-length parameter to be chosen such that the monotone Armijo condition
\begin{equation}J(\ve x^{(k)} +\lambda_k\ve d^{(k)}) \leq J(\ve x^{(k)}) + \beta \lambda_k\nabla J(\ve x^{(k)})^T\ve d^{(k)} \label{Armijo}\end{equation}
is satisfied for a fixed value of the parameter $\beta \in (0,1)$, in order to guarantee the sufficient decrease of the objective function. In practice, $\lambda_k$ is computed by a standard backtracking condition as $\lambda_k = \theta^{m}$, where $\theta\in (0,1)$ and $m$ is the smallest integer such that \eqref{Armijo} is satisfied.

The convergence of the SGP scheme, which is outlined in Algorithm \ref{GPM}, 
can be proved when the step-size $\alpha_k$ and the diagonal entries of $D_k$ 
are bounded above and away from zero, i.e. 
$\alpha_k\in [\alpha_{min},\alpha_{max}]$ with 
$0<\alpha_{min}<\alpha_{max}$ and $D_k$ is chosen in the set ${\mathcal D} $ of 
diagonal matrices whose diagonal entries have values between $L_1$ and $L_2$, 
for given thresholds $0<L_1< L_2$.

\begin{algorithm}[ht]
\caption{Scaled gradient projection (SGP) method}
\label{GPM}
Choose the starting point $\ve x^{(0)} \geq 0$ and set the parameters $\beta, \theta\in (0,1)$,
$0< \alpha_{min} <\alpha_{max}$.\\[.2cm]
{\textsc{For}} $k=0,1,2,...$ \textsc{do the following steps:}
\begin{itemize}
\item[]
\begin{AlgorithmSteps}[4]
\item[1] Choose the parameter $\alpha_k \in [\alpha_{min},\alpha_{max}]$ and the scaling matrix $D_k\in \mathcal D$;
\item[2] Compute the descent direction:\\ $\ve d^{(k)}=P_{\Omega,D_k^{-1}}(\ve x^{(k)}-\alpha_kD_k\nabla J(\ve x^{(k)})) -\ve x^{(k)}$;
\item[3] Backtracking loop: compute the smallest integer $m$ such that \eqref{Armijo} is satisfied with $\lambda_k=\theta^{m}$;
\item[4] Set $\ve x^{(k+1)} = \ve x^{(k)} + \lambda_k  \ve d^{(k)}$.
\end{AlgorithmSteps}
\end{itemize}
\textsc{End}
\end{algorithm}
The SGP algorithm has been recently applied in several image
restoration problems (see e.g. \cite{benvenuto2010,bonettini2010,zanella}).
Under standard assumptions, it can be proved \cite{bonettini2009} that the SGP 
algorithm is well defined and any limit point of the sequence $\{\xk\}$ is a 
stationary point of \eqref{minprob}; if, in addition, $J(\ve x)$ is convex, any 
limit point is a minimum point. It is worth stressing that the convergence 
result holds for any choice of the scaling matrix 
$D_k\in {\mathcal D}$ and the step-length $\alpha_k\in[\alpha_{min},\alpha_{max}]$: this 
freedom of choice can be exploited in order to improve the convergence speed.
Indeed, it is well known that gradient methods can be significantly accelerated 
by a clever choice of the step-length parameter $\alpha_k$: one of the more 
effective strategies are the Barzilai--Borwein (BB) rules proposed firstly in 
\cite{Barzilai-Borwein-1988} for quadratic unconstrained programming and 
then developed and analyzed for more general problems 
(see \cite{Dai-Fletcher-2005,Dai-etal-2006,Frassoldati-etal-2008,Zhou-etal-2006} 
and reference therein). The BB rules can be considered a very cheap way to 
capture the second order information enforcing a quasi-Newton property. In our 
algorithm we adopt the scaled versions of the BB rules proposed in 
\cite{bonettini2009}, which are given by
\begin{equation}
\label{BB1-2}
\alpha_k^{{(BB1)}} = \frac{{\ve s^{(k-1)}}^T D_k^{-1} D_k^{-1}
{\ve s}^{(k-1)}}{{{\ve s}^{(k-1)}}^T D_k^{-1} {\ve z}^{(k-1)}}\,, \qquad
\alpha_k^{{(BB2)}} = \frac{{{\ve s}^{(k-1)}}^T D_k {\ve z}^{(k-1)} }
{{{\ve z}^{(k-1)}}^T D_k D_k {\ve z}^{(k-1)}}\,,
\end{equation}
where ${\ve s}^{(k-1)}={\ve x}^{(k)}- {\ve x}^{(k-1)}$ and
${\ve z}^{(k-1)}=\nabla J({ \ve x^{(k)}})-\nabla J({ \ve x^{(k-1)}})$. Based on the previous formulas, we define the values $\alpha_k^{(1)},\alpha_k^{(2)}\in [\alpha_{min},\alpha_{max}]$ in the following way
\vskip0.2cm
\noindent
\hspace*{0.3cm}
\textsc{if} $ {{\ve s^{(k-1)}}}^T D_k^{-1} {\ve z}^{(k-1)} \le 0$ \textsc{then}\\
\hspace*{.5cm} $\alpha_k^{(1)} = \min\left\{10 \cdot \alpha_{k-1}, \ \alpha_{max} \right\}$; \\
\hspace*{.3cm}
\textsc{else}\\
\hspace*{.5cm} $\alpha_k^{(1)} = \min\left\{\alpha_{max}, \ \max\left\{ \alpha_{min}, \ \alpha_k^{{(BB1)}} \right\}\right\}$; \\
\hspace*{.3cm}
\textsc{endif} \\
\hspace*{0.3cm}
\textsc{if} $ {{\ve s^{(k-1)}}}^T D_k {\ve z}^{(k-1)} \le 0$ \textsc{then}\\
\hspace*{.5cm} $\alpha_k^{(2)} = \min\left\{10 \cdot \alpha_{k-1}, \ \alpha_{max} \right\}$; \\
\hspace*{.3cm}
\textsc{else}\\
\hspace*{.5cm} $\alpha_k^{(2)} = \min\left\{\alpha_{max}, \ \max\left\{ \alpha_{min}, \ \alpha_k^{{(BB2)}} \right\}\right\}$; \\
\hspace*{.3cm}
\textsc{endif}
\\[.1cm]
From the numerical experience, the best performances are obtained by an alternation of the two BB formulas: thus, following \cite{Frassoldati-etal-2008}, we choose the following criterion for computing $\alpha_k$
\vskip0.2cm
\noindent
\hspace*{0.3cm}
\textsc{if} $k\leq 20$ \textsc{then}
\begin{equation} \hspace*{.6cm} \alpha_k = \min_{j=\max\left\{1,k+1-M_{\alpha}\right\},\dots,k} \ \alpha_j^{(2)};\label{alphadef}\end{equation}
\hspace*{.3cm}\textsc{else if} $ {\alpha_k^{(2)}}/{\alpha_k^{(1)}} \le
\tau_k$ \textsc{then}\\
\hspace*{.5cm} {Set} $\alpha_k$ as in \eqref{alphadef} \\
\hspace*{.5cm} $\tau_{k+1} = 0.9 \cdot \tau_{k}$;\\
\hspace*{.3cm}
\textsc{else}\\
\hspace*{.5cm} $\alpha_k = \alpha_k^{(1)}; \qquad \tau_{k+1} = 1.1 \cdot \tau_{k}$;\\
\hspace*{.3cm}
\textsc{endif}
\\[.1cm]
where $M_{\alpha}$ is a prefixed positive integer and $\tau_1\in (0,1)$. Our 
choice of taking the value defined in \eqref{alphadef} for the first 20 
iterations leads to a more stable behaviour and, in some cases, also to a 
slight improvement of the reconstruction accuracy \cite{Prato-etal-2012}.

The other crucial ingredient for the practical performances of SGP is the 
choice of the scaling matrix: in our case, taking into account the 
objective functions of \eqref{probf} and \eqref{probh}, we adopt the scaling 
suggested by the RL algorithm
\begin{equation*}
D_k = \mbox{diag}\left(\min(L_2,\max(L_1,\xk)\right)
\end{equation*}
as suggested also in \cite{bonettini2009}. Obviously, in the inner
iterations of (\ref{probf}), $\ve x$ is replaced by $\ve f$ while, in the
inner iterations of (\ref{probh}), $\ve x$ is replaced by $\ve h$.
As concerns the choice of the bounds $(L_1,L_2)$, at the beginning of each 
inner subproblem we perform one step of the RL method and tune the parameters
according to the min/max positive values $y_{\min}$/$y_{\max}$
of the resulting image according to the rule
\vskip0.2cm
\noindent
\hspace*{0.3cm}
\textsc{if} $ y_{\max}/y_{\min} < 50$ \textsc{then}\\
\hspace*{.5cm} $L_1 = y_{\min}/10$;\\
\hspace*{.5cm} $L_2 = y_{\max} \cdot 10$;\\
\hspace*{.3cm}
\textsc{else}\\
\hspace*{.5cm} $L_1 = y_{\min}$;\\
\hspace*{.5cm} $L_2 = y_{\max}$;\\
\hspace*{.3cm}
\textsc{endif}

\subsection{Computing the projections}

Since the minimization steps \eqref{probf} and \eqref{probh} involve 
different constraints, corresponding to two convex sets $\Omega_1$ and 
$\Omega_2$, respectively, we have to account for two different algorithms 
to compute the projections $P_{\Omega_1,D_k^{-1}}$ and $P_{\Omega_2,D_k^{-1}}$.
In the alternating procedure of Algorithm \ref{CSGP}, when SGP is applied 
to problem \eqref{probf}, the projection consists of a simple component 
thresholding, obtained by setting all the negative elements of the vector 
to be projected equal to zero. For the updating of the PSF, instead, we have 
to project on the constraints set of the problem \eqref{probh}, consisting of 
a single linear equality constraint, in addition to simple bounds 
(box constraints) on the variables. The resulting constrained optimization 
problem to be addressed is therefore
\begin{eqnarray}
\label{projprobl}
& &{\rm min}~~  \frac{1}{2} \ve z^T D_k^{-1} 
\ve z - \ve z^T \ve y \\ \nonumber
& &{\rm s.t.}~~~0\leq \ve z\leq s ~,~\sum_{\ve i\in S} \ve z_{\ve i} = 1
\end{eqnarray}
where $\ve y = D_k^{-1}(\ve x^{(k)}-\alpha_kD_k\nabla J(\ve x^{(k)}))$.
By introducing the Lagrangian penalty function, one can see that the 
orthogonal projection \eqref{projprobl} can be re-conducted to a root-finding 
problem of the piecewise linear monotonically non-decreasing function
\begin{equation*}
t(\xi) = \sum_{\ve i \in S} \ve z_{\ve i}(\xi) - 1 = 0,
\end{equation*}
where $\xi$ is the Lagrangian multiplier of the equality constraint, 
\begin{equation*}
\ve z_{\ve i}(\xi) = {\rm{mid}}(0,(D_k)_{ii}(\ve y_{\ve i} + \xi),s)
\end{equation*}
and ${\rm{mid}}(a_1, a_2, a_3)$ is the component-wise
operation that supplies the median of its three arguments.
For solving this kind of problem we apply the secant-based method proposed in 
\cite{Dai-Fletcher-2006} (its Matlab implementation is given in the 
corresponding technical report downloadable at the webpage 
http://www.maths.dundee.ac.uk/nasc/na-reports/NA216\_RF.pdf), which is able 
to compute the projection very quickly and whose computational cost grows 
linearly in time with respect to the image size \cite[\S 3.1]{bonettini2009}.

\section{Numerical experiments}

In this section we investigate the effectiveness of the proposed blind method 
by means of several numerical experiments. Since the blind problem 
formulated in (\ref{prob}) is non-convex, 
several local minima may exist. Moreover, we know that any limit point 
(or the limit) of the proposed iteration is a stationary point of the 
problem. The limit depends, in general, on the numbers of inner
iterations but also on the initialization of the outer iteration; therefore
it is important to initialize the procedure with a sensible initial guess
and we first discuss this point.

As concerns the object we can use the standard initialization of the RL 
algorithm, namely a constant object with a flux coinciding with the flux of
the image after background subtraction. The choice of the initial PSF 
is more important because, in the first step of the procedure, the image is 
deconvolved with this PSF. 

To this purpose we point out an important
property of the PSF of a telescope: it is a band-limited function and,
if the telescope consists of a circular mirror, the band, i.e. the support
of its Fourier transform, is a disc with a radius proportional to the ratio
between the diameter $D$ of the telescope and the observation wavelength
$\lambda$. It is not easy to insert this property as a constraint on the PSF
because the projection on the resulting set of constraints (including
SR and normalization) is not easily computable. For this reason we do not
consider this constraint in this paper. However we can try to force the 
estimated PSF to have this property using an initial PSF which is 
band-limited and satisfies the other constraints. 

The ideal PSF of the telescope is not suitable as initial guess because it does 
not satisfy the SR constraint. However one can consider, as suggested for instance 
in \cite{biggs}, the autocorrelation of the ideal PSF, which has the same band. 
In our simulations, which assume in general a telescope of the 8m class and observations 
in H-band (see the beginning of the next section), the resolution, inversely proportional 
to the bandwidth, is about 50 mas (milliarcseconds). Using an oversampling such that the 
pixel size is 15 mas, the diameter of the band in Fourier space is about 186 pixels 
(remark that it depends also on the number of pixels in the image). With these values, 
the autocorrelation is a good choice if SR $\geq$ 0.46 (the value of $s$ depends on both 
SR and the ratio $D/\lambda$); for lower values of SR one can take the autocorrelation 
of the autocorrelation and so on, until the SR constraint is satisfied. This is the choice 
considered in our numerical experiments and, quite surprisingly, it seems that the algorithm, 
in spite of its high nonlinearity, preserves the band-limiting property satisfied by the initial guess. 

All the numerical experiments have been performed with a set of routines 
implemented by ourselves in Interactive Data Language (IDL). The codes of 
the algorithms presented and discussed in this paper are available under 
request.

\subsection{Image generation}

As mentioned in the Introduction the use of SR as a constraint on the PSF is
first proposed in \cite{desidera2009}. Therefore some of our numerical
experiments coincide with some of the tests performed in that paper. In 
particular we use three of the AO-corrected PSFs (with SR equal to 0.67, 
0.40 and 0.17, respectively), used by these authors and obtained by means of 
the Software Package CAOS \cite{carbillet2004a}; the parameters corresponding 
to these PSFs are given in \cite{desidera2009}. We only specify that they
correspond to a telescope with an effective diameter of 8.22 m and an 
observation wavelength of $\lambda=1.65 \mu$m (H-band). For each PSF, the 
images are generated by assuming, as in \cite{desidera2009} a time exposure 
of 1200 s, with a total transmission of 0.3. Moreover, a background of 13.5 
mag arcsec$^{-2}$, corresponding to observations in H-band, is added to the 
blurred images (for the convenience of the reader we remark that it 
corresponds to $3.41 \times 10^4$ counts per pixel). The results are 
perturbed with Poisson noise and additive Gaussian noise with 
$\sigma=10~e^-/$px. According to the approach proposed in \cite{snyder1995} 
and discussed in Sect. 2, RON compensation is obtained in the deconvolution 
algorithms by adding the constant $\sigma^2=100$ to the images and the 
background. 

In our first experiment we also consider an example which is not related to 
AO imaging but is a simulation of HST image before COSTAR correction,
since this image is frequently used in the testing of deconvolution methods.
Obviously in such a case the ideal PSF must be computed, by taking into
account the diameter of the Hubble telescope, about 2.4 m, and the
assumed observation wavelength of about 0.55 $\mu$m, and compared to the
aberrated PSF in order to estimate the corresponding SR. 
Objects, PSFs and blurred images used in all our experiments are sized 
$256 \times 256$ pixels.

\subsection{Point-wise objects}

We first report results on the following examples:
\begin{itemize}
\item the binary system considered by Desider\`a \& Carbillet 
\cite{desidera2009}, in which the two components have the same magnitude 12 
in H-band (corresponding approximately to $6.03 \times 10^8$ counts) 
with an angular separation of 285 mas (19 pixels), i.e. 
$\sim$7 times larger than the diffraction limit ($\sim$40 mas);
\item a model of an open star cluster based on an image of the Pleiades, 
consisting of 9 stars with magnitudes ranging from about 13 (i.e. about
$2.32 \times 10^8$ counts) to 16 (about $1.79 \times 10^7$ counts) in H-band
and described in \cite{bertero2011};
\item a simulation of a star cluster, consisting of 470 light sources, 
as observed by the Hubble Space Telescope (HST) before COSTAR correction.
For this case only, we do not use an AO-corrected PSF but the aberrated
HST PSF, which corresponds to SR=0.09.
These data can be obtained via anonymous ftp from \\ 
ftp://ftp.stsci.edu/software/stsdas/testdata/restore/sims/star\_cluster/. 
\end{itemize}
In Fig. 1 we show the images of the binary and of the star 
cluster in the case of a PSF with SR=0.67 as well as the HST image of a
simulated star cluster.
\begin{figure}
\begin{center}
\includegraphics[width=0.7\textwidth]{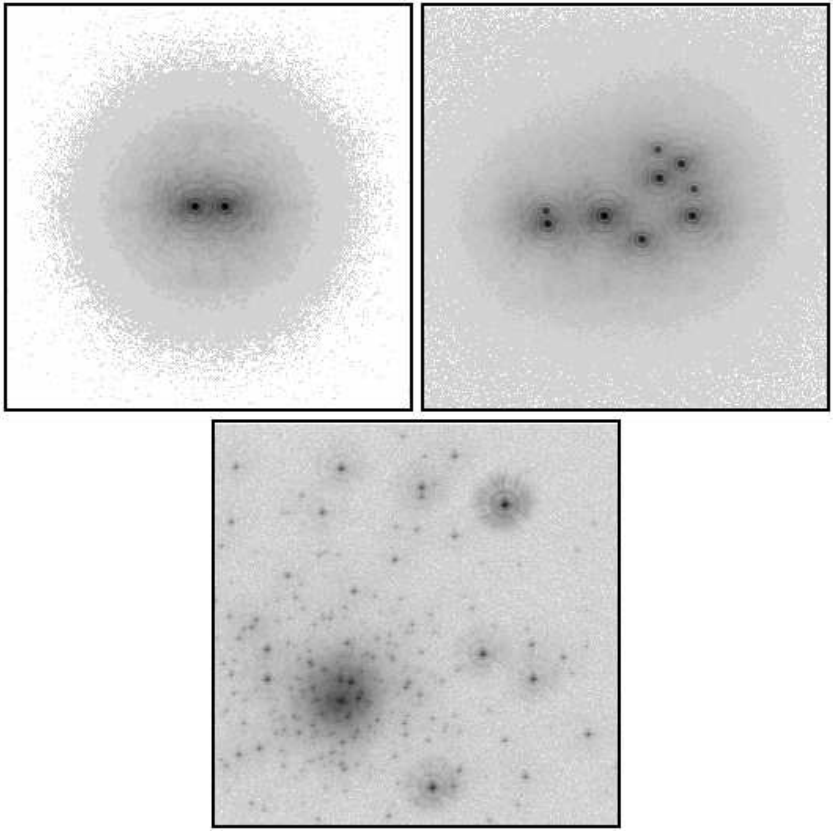}
\label{fig1.eps}
\end{center}
\caption{Images of the binary and of the star cluster (upper panels) in
the case of a PSF with SR=0.67. In the lower panel the image of the HST
star cluster.}
\end{figure}
For all these examples we use 50 inner iterations on the object and 1
inner iteration for the PSF. This choice can be justified by the
features of our blind problem discussed in the Introduction, since we need 
a sufficiently large number of SGP inner iterations for obtaining a nearly
point-wise object. Moreover, with a few experiments on the binary, we verify 
that this choice is a good compromise which provides a sufficiently fast 
convergence for all cases. In a first instance we perform 300 outer 
iterations. 

As concerns the measure of the quality of the reconstructions, for the PSFs 
we use the relative r.m.s. error between the reconstructed PSF $\ve h$ and 
that used for image generation $\ve {\tilde h}$, i.e.
\begin{equation}
\label{rmse}
RMSE=\frac{||{\ve h} - {\ve {\tilde h}}||_2}{||\ve {\tilde h}||_2}~~,
\end{equation}
where $||.||_2$ denotes the $\ell_2$ norm. The same parameter is used for 
measuring the quality of the reconstruction of the HST star cluster.
In the case of the binary and of the open star cluster we use a 
{\it magnitude average relative error} (MARE) defined as follows
\begin{equation}
\label{mare}
MARE=\frac{1}{q}\sum_{i=1}^q\frac{|m_i-\tilde m_i|}{\tilde m_i}~~,
\end{equation}
where $q$ is the number of stars and $m_i$, $\tilde m_i$ are respectively 
the reconstructed and the true magnitudes.  

\begin{table}
\caption{Reconstruction errors for point-wise objects. In the first
column we specify the object and in the second the value of the SR used
for image generation; in the third and fourth the values of MARE (RMSE 
in the case of HST cluster) when SGP is used for image deconvolution
with the exact and initial PSF respectively. In the fifth column   
the values of MARE (RMSE in the case of HST) obtained with 300 outer 
iterations by our blind approach, using the fixed pair 
$(n_f,n_h) = (50,1)$. Finally, in the 
last two columns the errors between the true and the initial PSF, 
followed by the errors between the true PSF and that provided by the 
blind approach.}
\label{tab2}
\begin{center}
\begin{tabular}{|c|c|c|c|c|c|c|}
\hline
Image & SR & $MARE$ & $MARE_1$ & $MARE_2$ & $RMSE_1$ & $RMSE_2$ \\

\hline \multirow{3}{*}{Binary}  
& 0.67 & $ 1.86 \times 10^{-5}$ & $ 1.12 \times 10^{-2}$ 
& $ 1.44 \times 10^{-3}$ & $32~\%$ & $1.8~\%$        \\
									
& 0.40 & $ 1.84 \times 10^{-5}$ & $ 2.26 \times 10^{-2}$
& $ 2.15 \times 10^{-3}$ & $54~\%$ & $2.9~\%$        \\
									
& 0.17 & $ 2.36 \times 10^{-6}$ & $ 1.67 \times 10^{-2}$ 
& $ 1.99 \times 10^{-3}$ & $55~\%$ & $3.3~\%$        \\

\hline \multirow{3}{*}{Cluster} 
& 0.67 & $ 2.10 \times 10^{-5}$ & $ 1.07 \times 10^{-2}$ 
& $ 3.09 \times 10^{-4}$ & $32~\%$ & $1.0~\%$  \\
									
& 0.40 & $ 4.43 \times 10^{-5}$ & $ 1.14 \times 10^{-2}$  
& $ 3.63 \times 10^{-4}$ & $54~\%$ & $1.1~\%$       \\
									
& 0.17 & $ 5.42 \times 10^{-5}$ & $ 1.30 \times 10^{-1}$   
& $ 2.87 \times 10^{-3}$ & $55~\%$ & $4.2~\%$        \\

\hline HST 						 
& 0.09 & $5.1~\%$ & $25~\%$ & $7.6~\%$  
& $47~\%$  & $6.7~\%$   \\

%
%
%
%
%
%
\hline
\end{tabular}
\end{center}
\end{table}

The results are shown in Table \ref{tab2} and are consistent with the
results reported in \cite{desidera2009} but obtained with a sound 
mathematical approach, allowing investigation of the limit for large
number of iterations and generalization to regularized problems. 
The values of MARE estimated
with our blind approach are certainly higher than those achievable if one 
deconvolves the data with the exact PSF ({\it inverse crime}) and given in
the third column, but they are still quite small. Moreover
the reconstruction of the PSFs is very satisfactory: the RMSEs of
the initial PSFs are of the order of 30-50 \%, while those of the
reconstructed PSFs are of the order of few percents. A comparison
between the true, initial and reconstructed PSFs is shown in Fig. 2.
We must add that the reconstruction error is still decreasing
after 300 iterations and therefore the minimum of the objective function is
not yet reached.

\begin{figure}
\begin{center}
\includegraphics[width=0.8\textwidth]{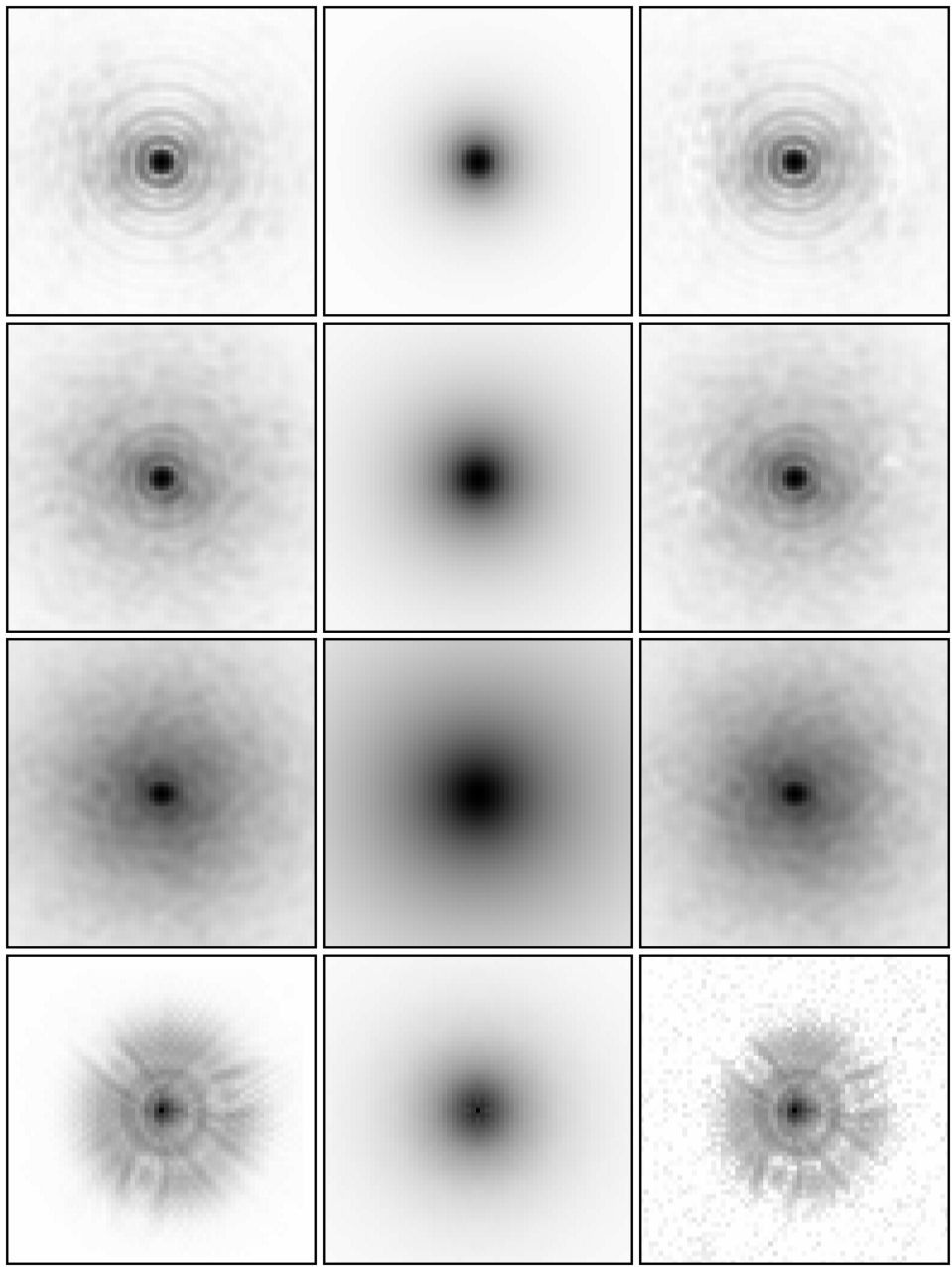}
\end{center}
\caption{First column: the PSFs used for image generation; second column: 
the PSFs used for initializing the blind algorithm (see the text for
their computation); third column: the PSFs reconstructed by the blind 
algorithm.
First row: AO-corrected PSF with SR=0.67; second row: AO-corrected PSF with
SR=0.40; third row: AO-corrected PSF with SR=0.17; fourth row: HST PSF
before COSTAR correction.}
\label{psf}
\end{figure}

For investigating the limit of the algorithm we consider three other 
examples of binaries: a binary with magnitudes 12-12 and angular distance 
of 10 pixels and two binaries with magnitudes 12-16 and angular distances 
of 19 and 10 pixels respectively. For these four examples of binaries we 
generate images using the PSF with the highest SR, namely 0.67, and we 
compute 8000 outer iterations of the blind algorithm, using the fixed pair 
$(n_f,n_h) = (50,1)$. In Table \ref{tab:binary067} we report the results 
obtained after 300, 4000 and 8000 outer iterations. We do
not find a uniform behaviour: in some cases the errors are decreasing for
increasing number of iterations while in others they are slightly 
increasing or do not have a monotone behaviour. In all cases the variations 
are small and the reconstruction errors on the PSF after 8000 iterations 
are quite small. However, in view of obtaining a sufficient accuracy, 300 
outer iterations can be sufficient in all cases. 

Similar results are 
obtained in the case of the open star cluster. Moreover, since the previous 
examples are derived from the examples considered in \cite{desidera2009}
and generated by assuming a very long observation time (hence a low noise
level) we also generated an image of the binary 12-12, angular distance
19 pixels, assuming an integration time of 12 seconds, with a reduction by
a factor 100 of the average number of photons. By performing 8000 iterations
we still find convergence of the algorithm but the error on the PSF is now
of the order of 1 \% and can be reached after 300 iterations. On the other
hand the value of MARE is quite satisfactory since it is of the order of 
$8 \times 10^{-4}$.   

\begin{table}
\caption{Reconstruction errors, provided by increasing number of iterations,
in the case of four different binaries (the parameters are indicated in the 
first column, as explained in the text) whose images are generated using 
the PSF with SR=0.67. As usual MARE is a measure of the errors on the 
magnitudes of the two stars while RMSE is a measure of the error on the 
reconstructed PSF.}
\begin{center}
\begin{tabular}{|c|c|c|c|c|}
\hline
\multicolumn{2}{|c|}{} & 300 it & 4000 it & 8000 it\\
\hline
12-12 & MARE & $1.44 \times 10^{-3}$  & $1.38 \times 10^{-4}$ & $1.31\times10^{-4}$ \\
\cline{2-5}
19 pixels & RMSE & $ 1.8~\%$ & $0.17~\%$& $ 0.15~\%$ \\
\hline
12-16 & MARE & $5.10 \times 10^{-4}$& $1.01 \times 10^{-3}$ & $1.52\times10^{-3}$\\
\cline{2-5}
19 pixels & RMSE & $0.99~\%$& $ 0.20~\%$& $ 0.27~\%$ \\
\hline
12-12 & MARE & $1.26 \times 10^{-3}$ & $1.42 \times 10^{-4}$ & $1.34 \times 10^{-4}$\\
\cline{2-5}
10 pixels & RMSE & $1.7~\%$& $0.18~\%$& $0.17~\%$ \\
\hline
12-16 & MARE & $1.31 \times 10^{-2}$& $1.92 \times 10^{-2}$& $2.22 \times 10^{-2}$\\
\cline{2-5}
10 pixels & RMSE & $1.1~\%$& $1.3~\%$ & $1.5~\%$\\
\hline
\end{tabular}
\label{tab:binary067}
\end{center}
\end{table}

It should be interesting to find a way for establishing if the minima we
find are the global ones or not, but, as it is known, global minimization
is a very difficult problem. As a test, even if it does not provide a 
proof that the minima are the global ones, we compare the minimum values of 
the objective function, i.e. the KL divergence, with its values corresponding
to the ground truths, i.e. the values obtained by substituting in Eq.
(\ref{KL_divergence}) the objects and PSFs used for image generation.
We find that the minimum values are of the order of $1.0 \times 10^4$
while the values corresponding to the ground truth are greater by about
a factor of 3. We can only say that, if these values were smaller than our
minimum values, then our minima were certainly local.

Before considering other examples it is important to remark that, in the 
case of the binary and of the small star cluster, the stars are 
reconstructed as single pixels with a sufficiently accurate flux value, but
the reconstructed images contain artifacts, in the sense that other
pixels take non zero values. These values are small but they can be
disturbing in the case of an accurate photometric analysis, for instance of
a star cluster, because they could be detected as faint stars. Indeed 
the difference of magnitude between the brightest artifact and the true 
stars is of the order of $\Delta m=8$. For this reason we perform an 
analysis of this problem in the case of the binary, using the PSF with the 
highest SR, namely SR=0.67.

\begin{figure}
\begin{center}
\begin{tabular}{ccc}
\fbox{\includegraphics[width=0.25\textwidth]{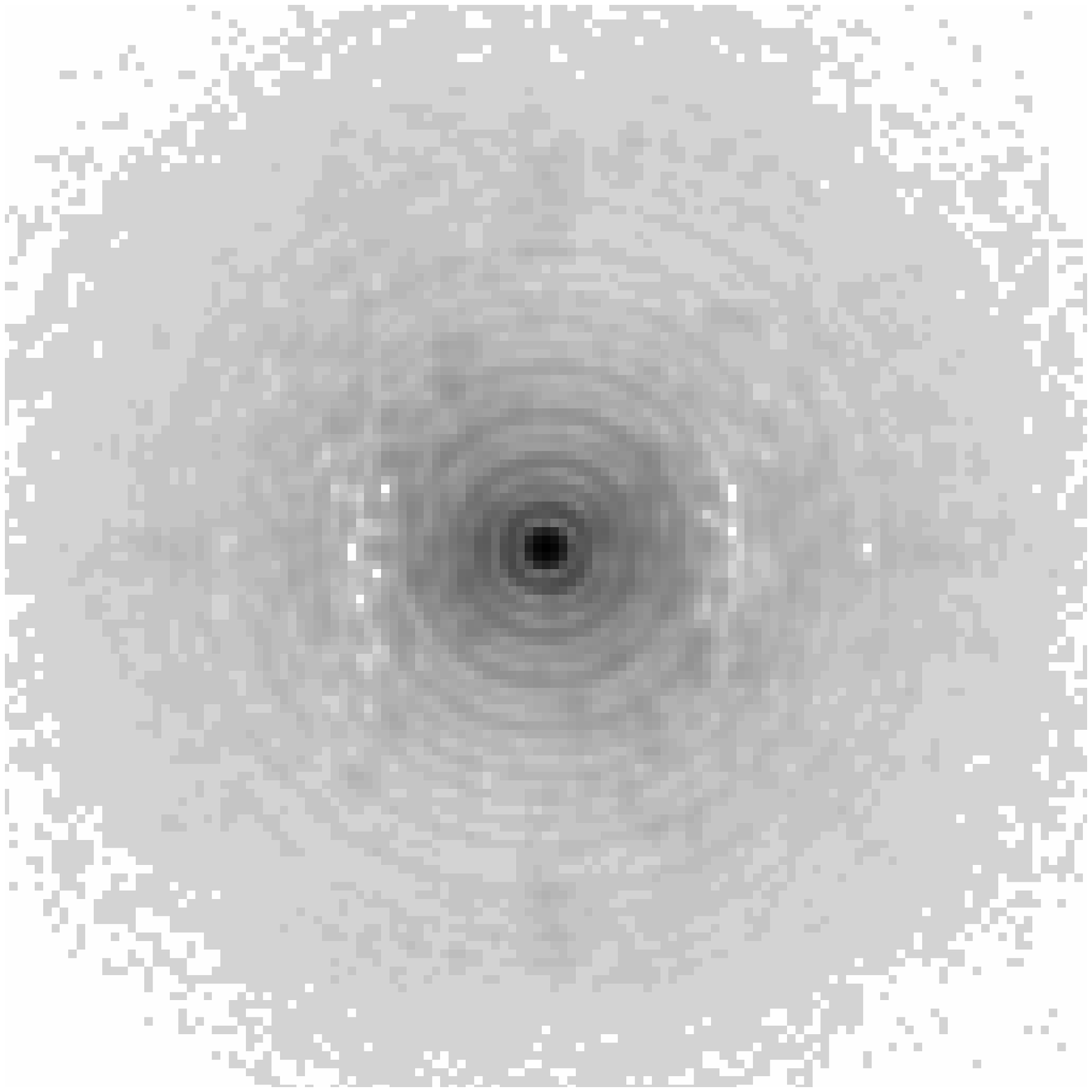}}&
\fbox{\includegraphics[width=0.25\textwidth]{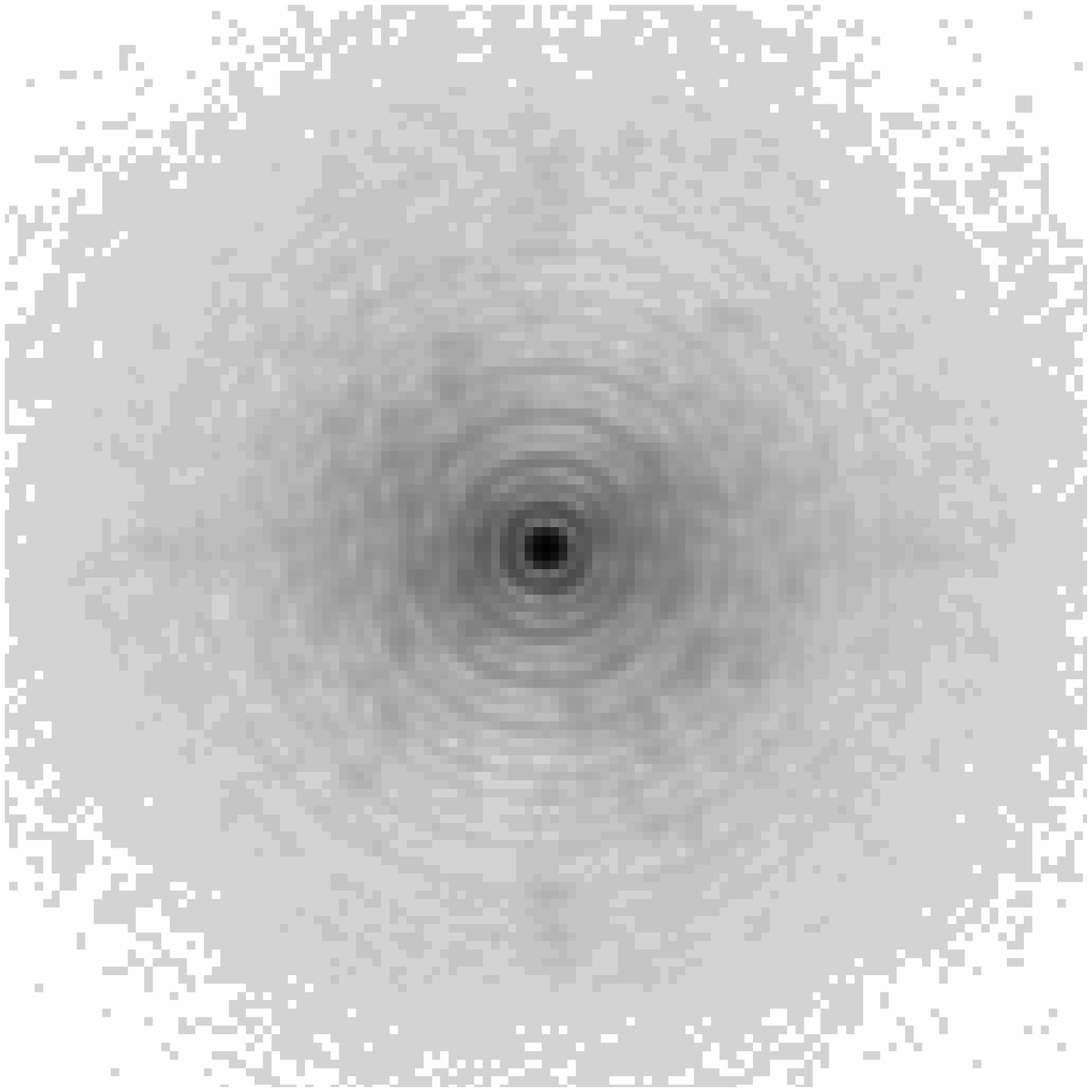}}&
\fbox{\includegraphics[width=0.25\textwidth]{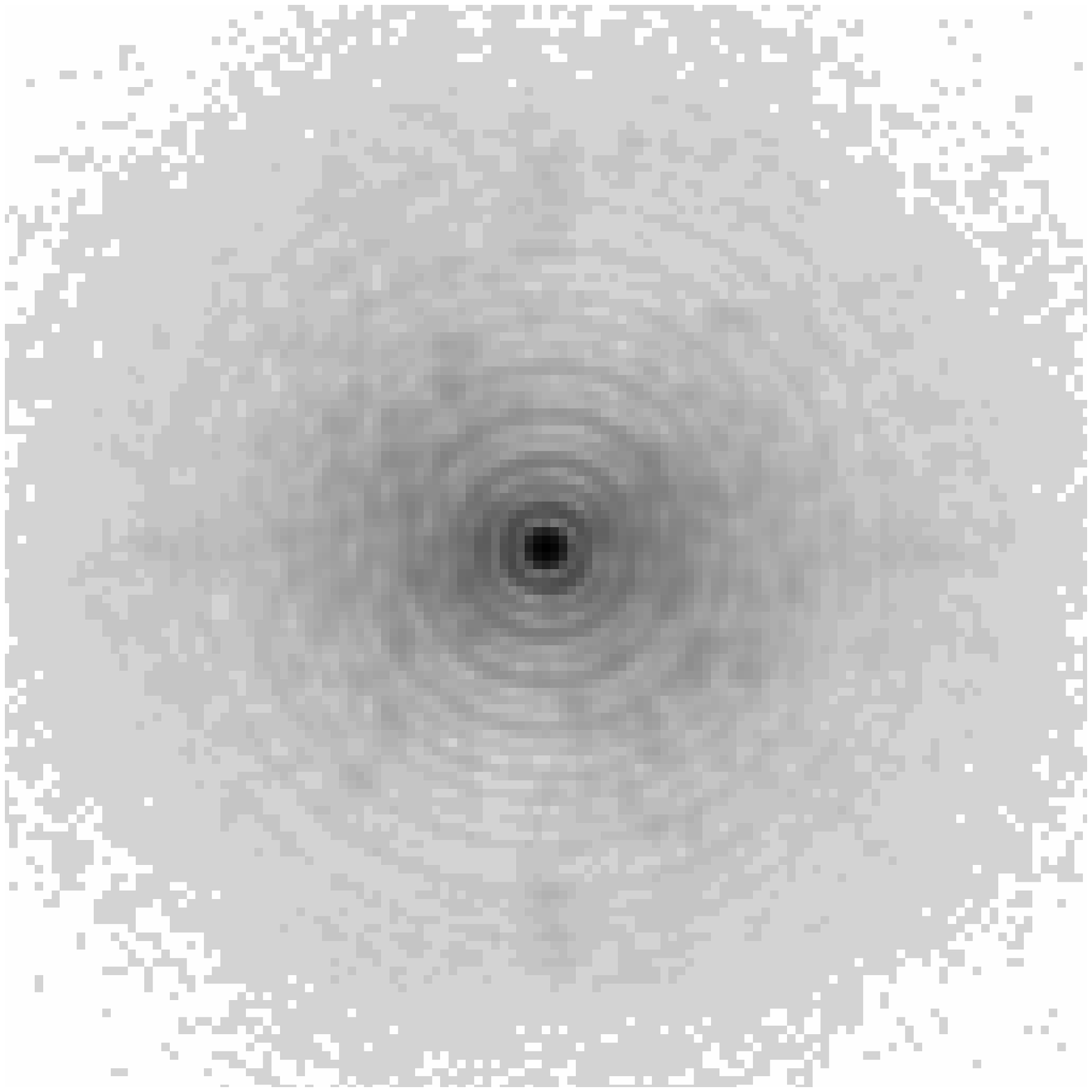}}\\
$RMSE = 1.8~\%$ & $RMSE = 0.17~\%$ & $RMSE = 0.15~\%$ \\
\fbox{\includegraphics[width=0.25\textwidth]{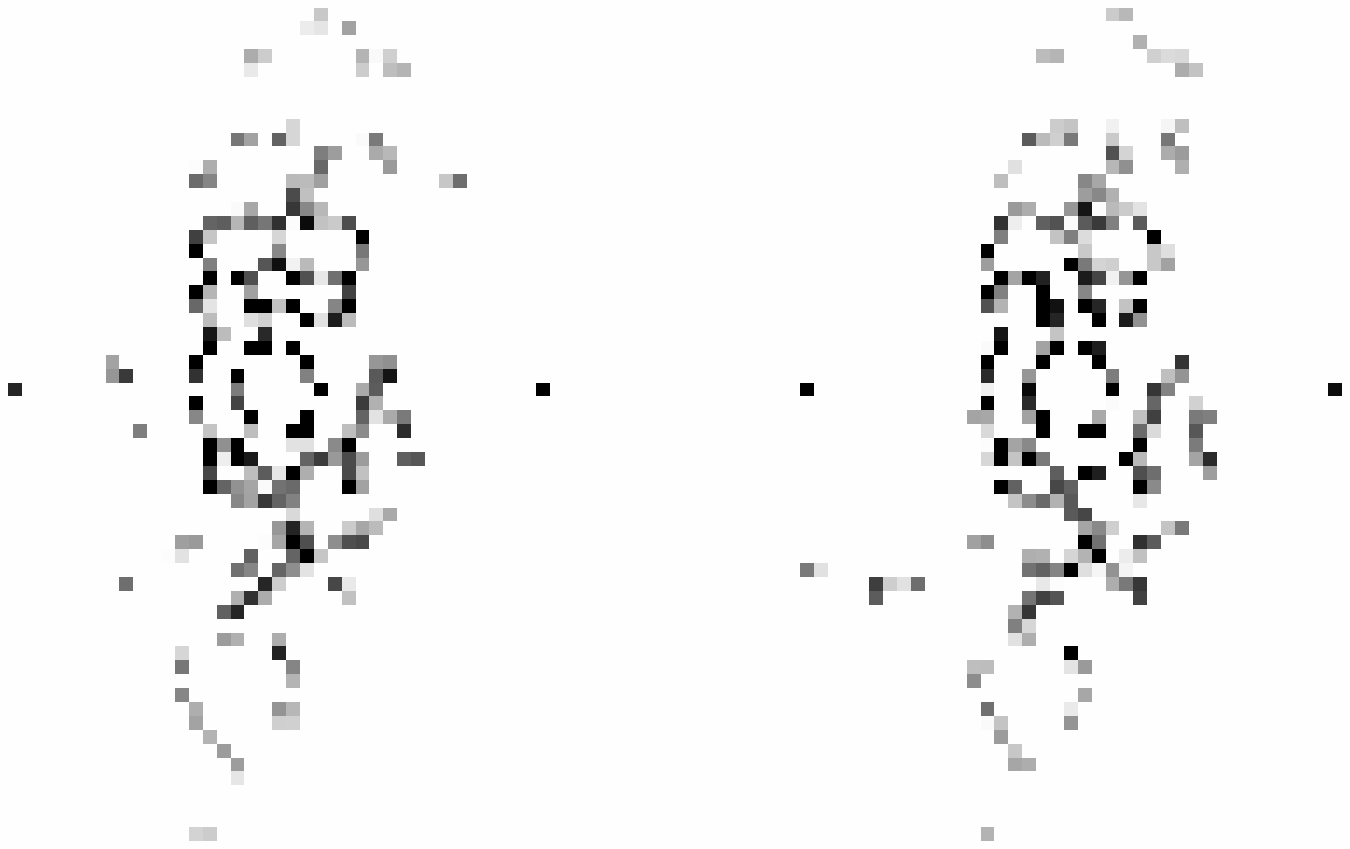}}&
\fbox{\includegraphics[width=0.25\textwidth]{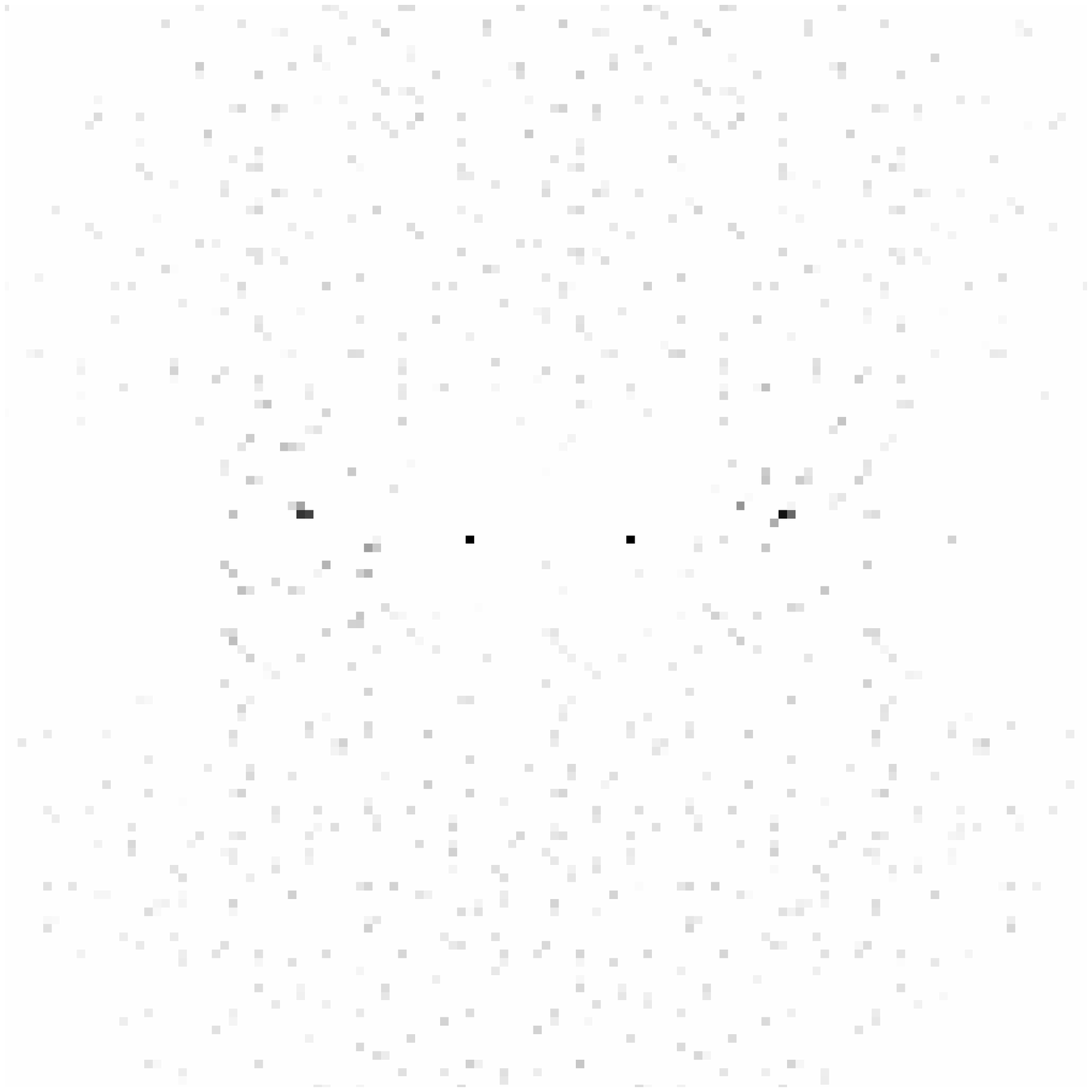}}&
\fbox{\includegraphics[width=0.25\textwidth]{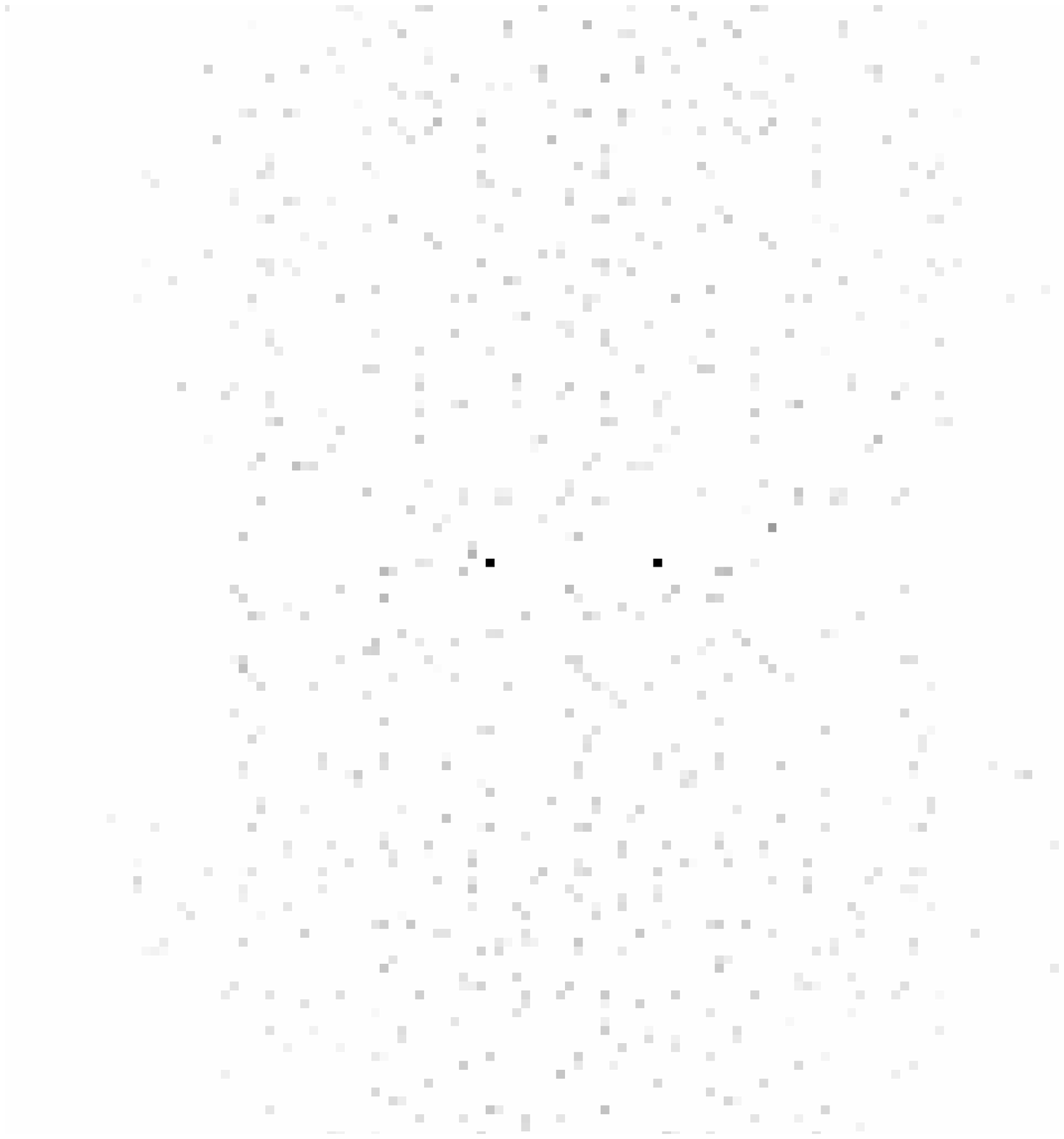}}\\
 $MARE = 1.44 \times 10^{-3}$ & $MARE = 1.38 \times 10^{-4}$ 
& $MARE = 1.31\times10^{-4}$ \\
\# = 524 & \# = 1653 & \# = 1480 \\
\fbox{\includegraphics[width=0.25\textwidth]{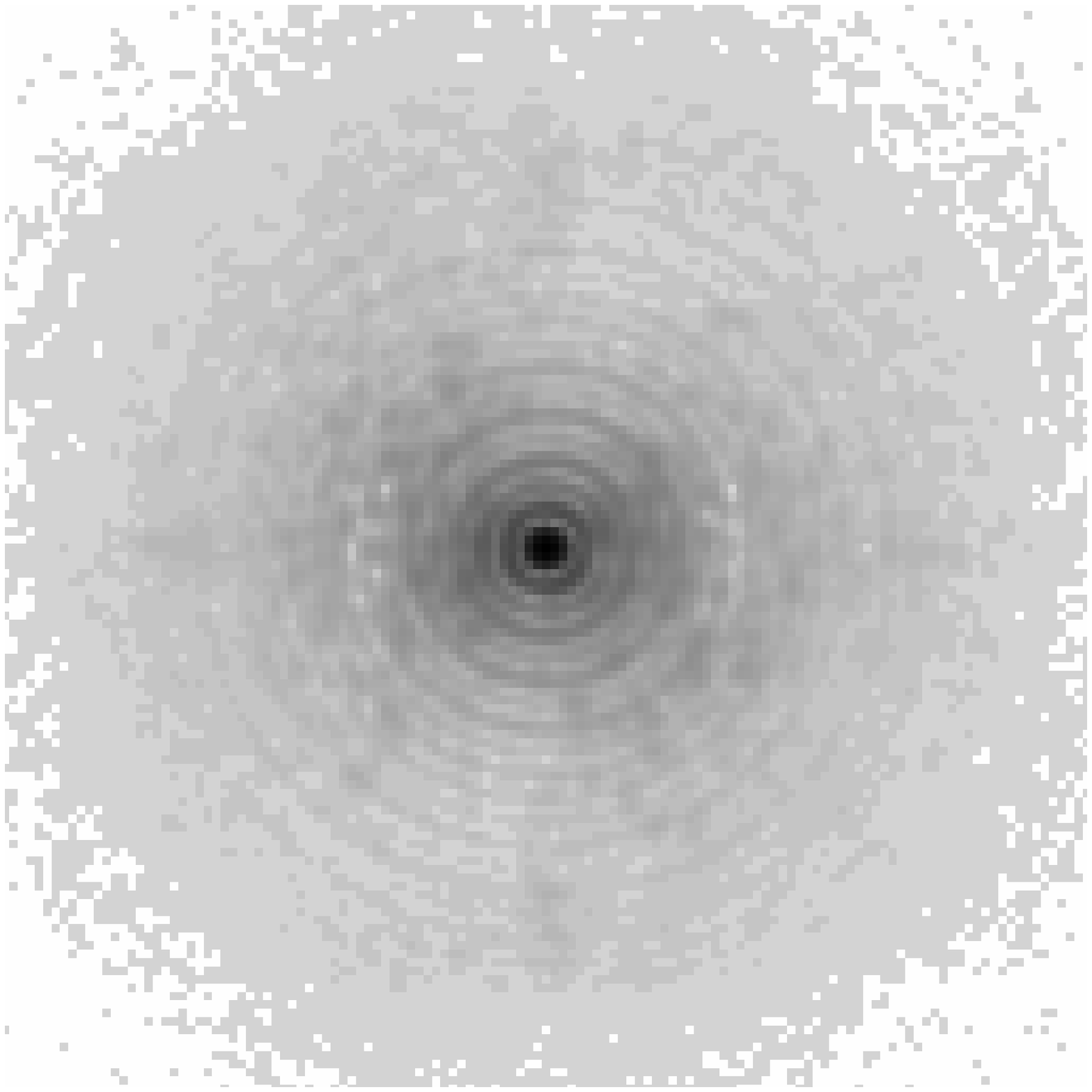}}&
\fbox{\includegraphics[width=0.25\textwidth]{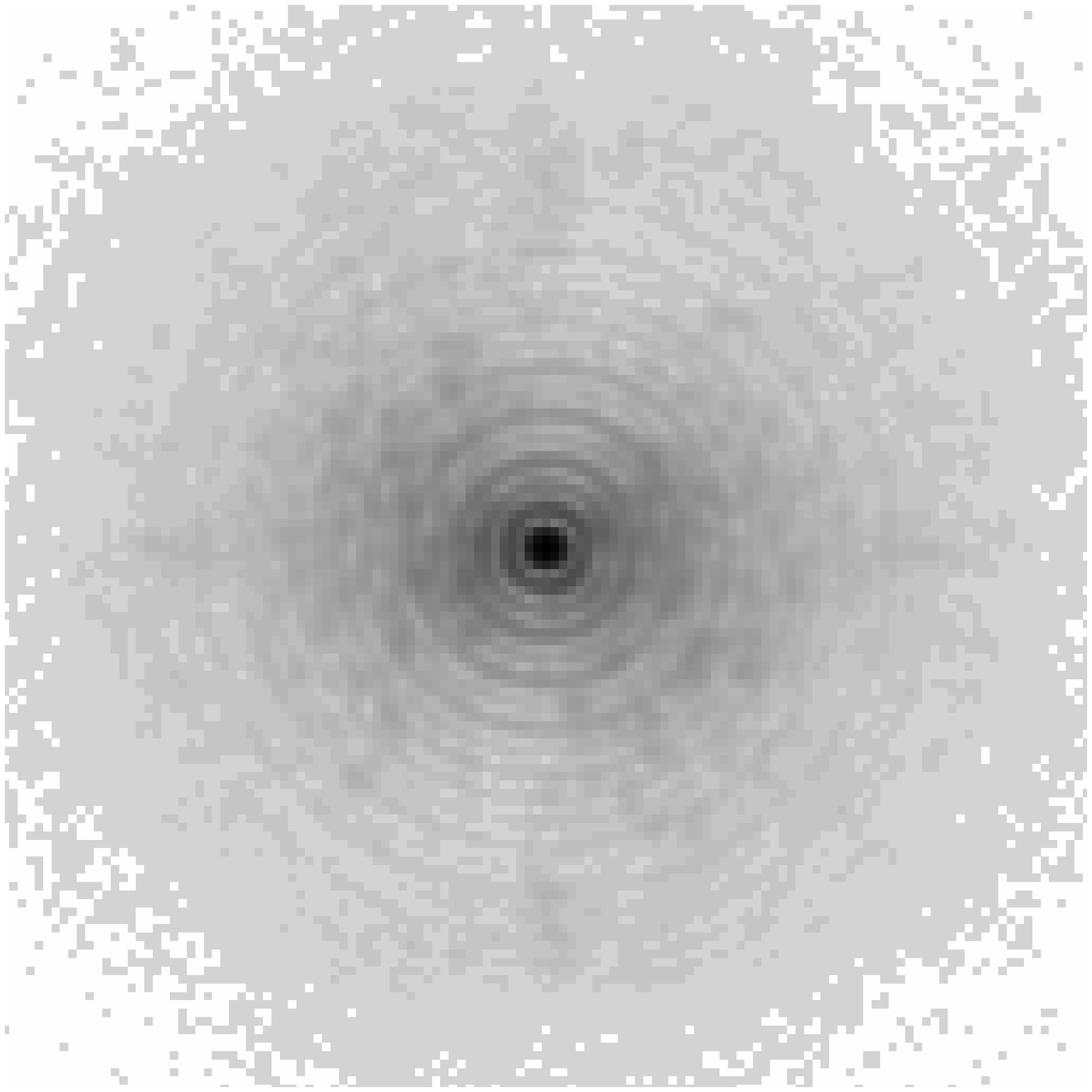}}&
\fbox{\includegraphics[width=0.25\textwidth]{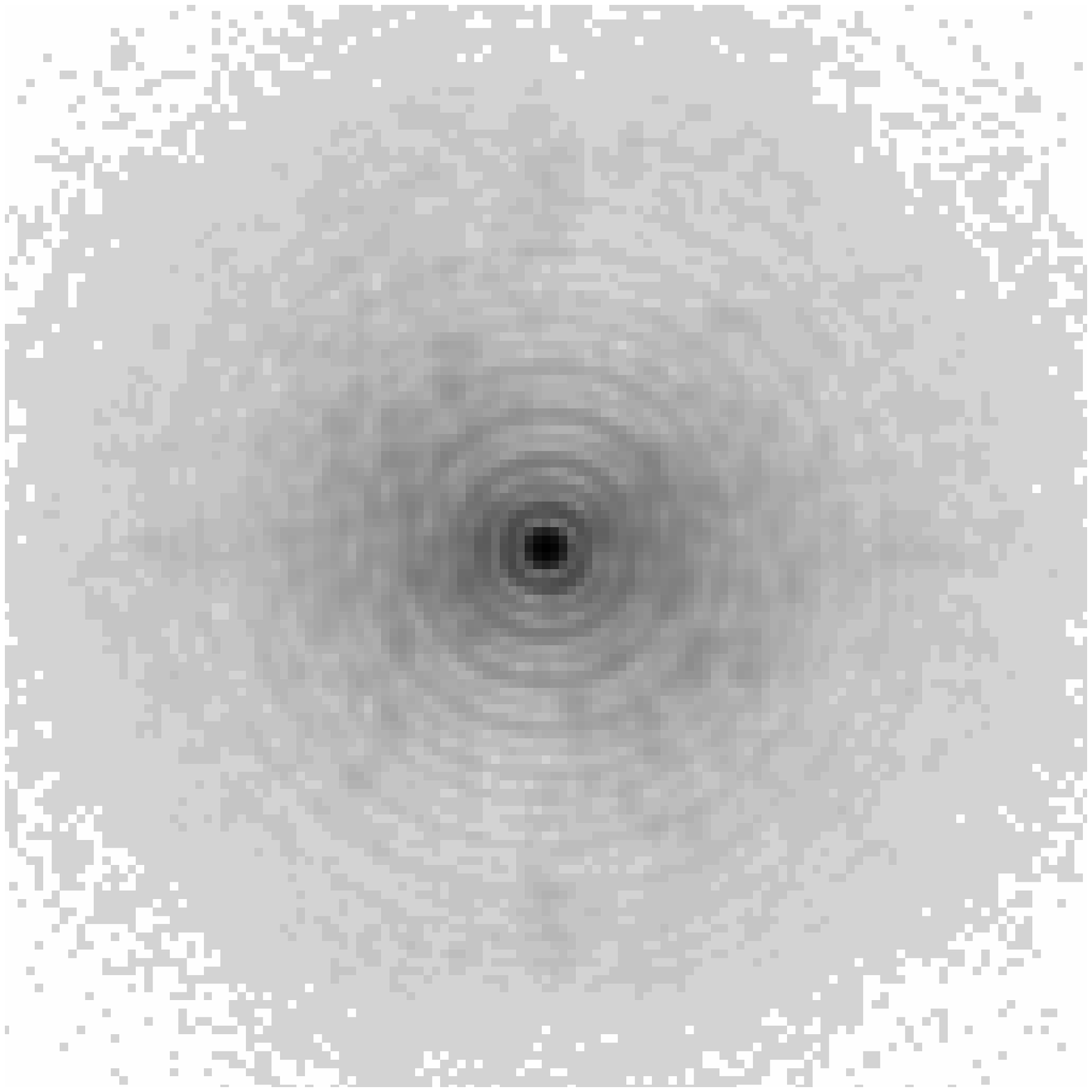}}\\
$RMSE = 2.1~\%$ & $RMSE = 2.0~\%$ & $RMSE = 2.0~\%$ \\
\fbox{\includegraphics[width=0.25\textwidth]{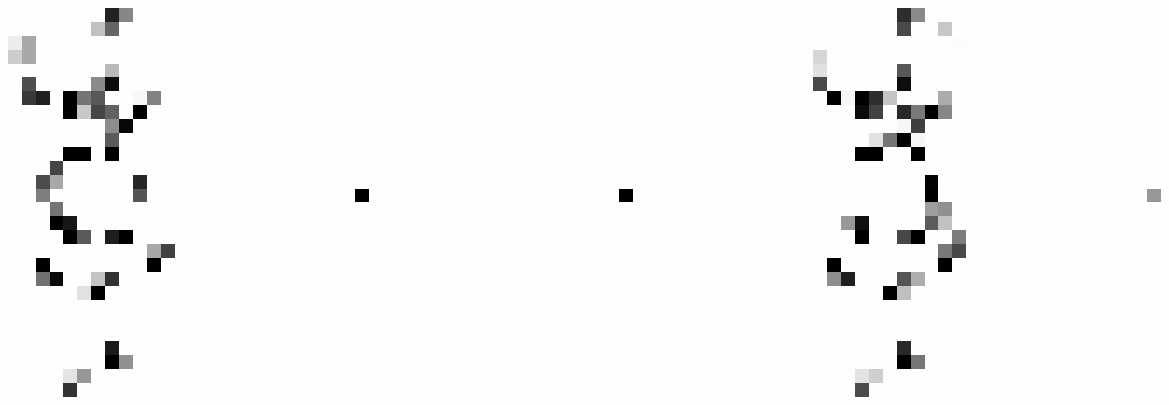}}&
\fbox{\includegraphics[width=0.25\textwidth]{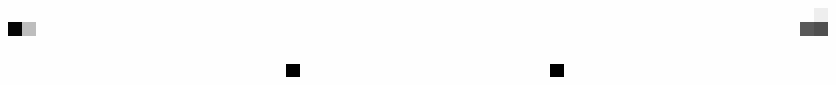}}&
\fbox{\includegraphics[width=0.25\textwidth]{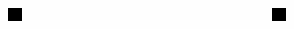}}\\
$MARE = 8.78 \times 10^{-5}$ & $MARE = 4.35 \times 10^{-4}$ & 
$MARE = 4.49\times10^{-4}$ \\
\# = 120  & \# = 5 & \# = 0 \\
\end{tabular}
\end{center}
\caption{Binary with magnitudes 12-12 and angular separation of 19 pixels. 
First and second row: the reconstructed PSF and object after 300 (left), 4000 
(middle), and 8000 (right) outer iterations with the exact value of SR 
(SR=0.67) as a constraint. Third and fourth rows: the reconstructed PSF 
and object after 300 (left), 500 (middle), and 2000 (right) outer 
iterations with the underestimated value of SR (SR=0.64) as a constraint. 
The symbol \# denotes the number of artifacts.}
\label{fig:artifacts}
\end{figure}

\subsection{Artifacts analysis}

We first consider the inverse crime reconstruction of the binary with
magnitudes 12-12 and angular distance 19 pixels. We deconvolve its image 
generated by means of the PSF with SR=0.67 using the same PSF (inverse
crime). The algorithm is the standard SGP with non-negativity constraint. 
Also in this case the reconstruction is not free of artifacts but they are 
randomly distributed and their values are quite small: the brightest 
artifact has a magnitude $m$ = 24, hence with a difference $\Delta m= 12$
with respect to the stars of the binary.

Next we apply the blind algorithm, using as a constraint the exact value
of SR and we analyze the results provided by the 8000 outer iterations 
already considered in the previous section. In the first two rows of
Fig. 3 we show the reconstructed PSF and the reconstructed object after
300, 4000 and 8000 outer iterations. The object is represented using a
logarithmic and gray-level reversed scale for stressing the artifacts.
After 300 iterations a few artifacts appear in the reconstructed PSF in 
a region corresponding to the positions of the two stars. Therefore, even 
if the reconstruction error is small, it is evident that it may be 
convenient to use a larger number of iterations. After 4000 iterations the 
PSF artifacts disappear and the reconstruction error is really very small, 
of the order of $0.2 \%$. The situation does not significantly change 
if we further increase the number of iterations. 

As concerns the reconstructed binary, after 300 iterations the artifacts
are concentrated along two arcs positioned around the two stars. We 
checked that this behaviour is stable if we change the noise realization.
Again, if we increase the number of iterations the results are better,
in the sense that the artifacts are more randomly distributed and the
intensity of the brightest one decreases. Indeed, after 4000 iterations
we have $\Delta m=10$ and after 8000 iterations $\Delta m=11$.
We find similar results in the case of the open star cluster and
therefore we can conclude that a very large number of iterations may
be required for obtaining very good results, at least if the exact
value of SR is known.

However, as briefly discussed in the Introduction, it is not possible to
know exactly the value of SR. According to astronomers the expected error
is of the order of 4 $\%$. Therefore, we consider a variation of the constraint
of this order of magnitude for images generated in the case of SR=0.67; more
precisely we consider two values, SR=$0.7$ and SR=$0.64$. We apply the blind
algorithm using as a constraint the corresponding values of $s$. In the first 
case the reconstructions are definitely worse, the number of artifacts 
considerably increases as well as the error on the PSF. For instance, in the 
case 12-12 the RMSE is of the order of 5 $\%$ and does not decrease with 
increasing number of iterations (remember that, in the case of exact value, 
the error after 8000 iterations is of the order of 0.15 $\%$). On the other 
hand, if we underestimate the SR, i.e. we take as a constraint the value of 
$s$ corresponding to SR=0.64, then the results are satisfactory. The 
reconstruction errors for the four binaries already considered in the previous 
subsection are reported in Table \ref{tab:binary064}. By comparing with the 
results reported in Table \ref{tab:binary067} and referring to the exact 
constraint, we can conclude that the reconstruction errors are not 
significantly greater than those obtained in the exact case and that the 
convergence is faster.

In the case of the binary with magnitudes 12-12 and angular distance 19
pixels we show the reconstructions of the PSF and of the binary after 300, 
500 and 2000 iterations respectively in the third and fourth row of Fig. 3. 
Quite surprisingly, the artifacts in the reconstruction of the binary 
completely disappear after 2000 iterations and the error on the
reconstructed PSF is of the order of 2 $\%$. We can add that the same result
is obtained in the case of the open star cluster. 

\begin{table}
\caption{Reconstruction errors in the case of the four different binaries of 
Table 2 (described in the text), considering an underestimated constraint 
of the blind algorithm (SR=0.64). As usual MARE is a measure of the errors on 
the magnitudes of the two stars while RMSE is a measure of the error on the 
reconstructed PSF.}
\begin{center}
\begin{tabular}{|c|c|c|c|c|}
\hline
\multicolumn{2}{|c|}{} & 300 it & 500 it & 2000 it\\
\hline
12-12 & MARE & $8.78 \times 10^{-5}$ & $4.35 \times 10^{-4}$& $4.49 \times 10^{-4}$ \\
\cline{2-5}
19 pixels & RMSE & $2.1~\%$& $2.0~\%$& $2.0~\%$\\
\hline
12-16 & MARE & $4.72 \times 10^{-4}$& $3.21 \times 10^{-3}$& $1.62\times10^{-1}$  \\
\cline{2-5}
19 pixels & RMSE &$2.0~\%$ & $2.1~\%$ & $4.0~\%$\\
\hline
12-12 & MARE & $4.36 \times 10^{-4}$  & $2.28 \times 10^{-4}$& $4.36\times10^{-4}$ \\
\cline{2-5}
10 pixels & RMSE & $2.4~\%$& $2.0~\%$& $2.0~\%$\\
\hline
12-16 & MARE & $1.99 \times 10^{-2}$& $5.91 \times 10^{-2}$ & ---\\
\cline{2-5}
10 pixels & RMSE & $2.4~\%$& $3.5~\%$& $4.0~\%$\\
\hline
\end{tabular}
\label{tab:binary064}
\end{center}
\end{table}

In order to further investigate the effect of a wrong value of $s$, in the
case of the binary 12-12 and SR=0.67, we also consider the cases
SR=0.4, 0.6, 0.8 and 1. The error on the PSF is of the order of 22 $\%$
in the case SR=0.4, about 5 $\%$ in the case SR=0.6 and about 11 $\%$ in
the two other cases. We can add that the artifacts disappear in the case of
underestimated SR. In conclusion an underestimate of SR of about 10 $\%$
(which does not correspond to the precision achievable in the experimental
estimation of this parameter) is still acceptable, while an overestimate can 
be dangerous in all cases.

\begin{table}
\caption{Reconstruction errors for complex and diffuse objects. In third and 
fourth columns, the best errors achieved by SGP with the true and the 
initial PSFs, respectively. In the fifth column, the best error 
obtained using a maximum of 100 outer iterations (for the choice of the
inner iterations see the text). Finally, in the last two columns, the error 
between the true PSF and the initial one, followed by the error 
between the true PSF and the one obtained in conjunction with the best 
reconstruction of the corresponding object.}
\label{tab:tab3}
\begin{center}
\begin{tabular}{|c|c|c|c|c|c|c|}
\hline
Image                   & SR   & $RMSE^{\, obj}$ & $RMSE_1^{\, obj}$ & $RMSE_2^{\, obj}$ & $RMSE_1^{\, psf}$ & $RMSE_2^{\, psf}$ \\
\hline
\multirow{3}{*}{Crab}   & 0.67 & $  11~\%$    & $  16~\%$    & $  12~\%$ & $ 32~\%$     & $ 6.7~\%$ \\
							          & 0.40 & $  12~\%$    & $  20~\%$    & $  14~\%$ & $ 54~\%$     & $  12~\%$ \\
								        & 0.17 & $  15~\%$    & $  22~\%$    & $  16~\%$ & $ 55~\%$     & $  16~\%$ \\
\hline
\multirow{3}{*}{Galaxy} & 0.67 & $  14~\%$    & $  23~\%$    & $  16~\%$ & $ 32~\%$     & $ 7.4~\%$ \\
												& 0.40 & $  16~\%$    & $  30~\%$    & $  19~\%$ & $ 54~\%$     & $  16~\%$ \\
												& 0.17 & $  20~\%$    & $  35~\%$    & $  23~\%$ & $ 55~\%$     & $  21~\%$ \\
\hline
\multirow{3}{*}{Nebula} & 0.67 & $ 3.2~\%$    & $ 6.8~\%$    & $ 6.8~\%$ & $ 32~\%$     & $  32~\%$ \\
												& 0.40 & $ 3.5~\%$    & $ 8.9~\%$    & $ 8.9~\%$ & $ 54~\%$     & $  53~\%$ \\
												& 0.17 & $ 4.2~\%$    & $ 9.0~\%$    & $ 7.9~\%$ & $ 55~\%$     & $  43~\%$ \\
\hline
\end{tabular}
\end{center}
\end{table}

\subsection{Complex and diffuse objects}

As additional examples of astronomical targets, we consider three HST images: 
the Crab nebula NGC 1952, the galaxy NGC 6946 and the planetary nebula NGC 7027. 
In all cases we assume an integrated magnitude equal to 10 and, for each one, 
we obtain three blurred images by convolving with the three PSFs of SR=0.67, 0.40
and 0.17. Again a background in H-band is added to all images and the 
results are perturbed with Poisson and additive Gaussian noise. In the first 
column of Fig. 4 we show the three objects in reversed scale of gray levels
while in the second column we show their blurred images in the case SR=0.67.

\begin{figure}
\begin{center}
\begin{tabular}{ccc}
\fbox{\includegraphics[width=0.25\textwidth]{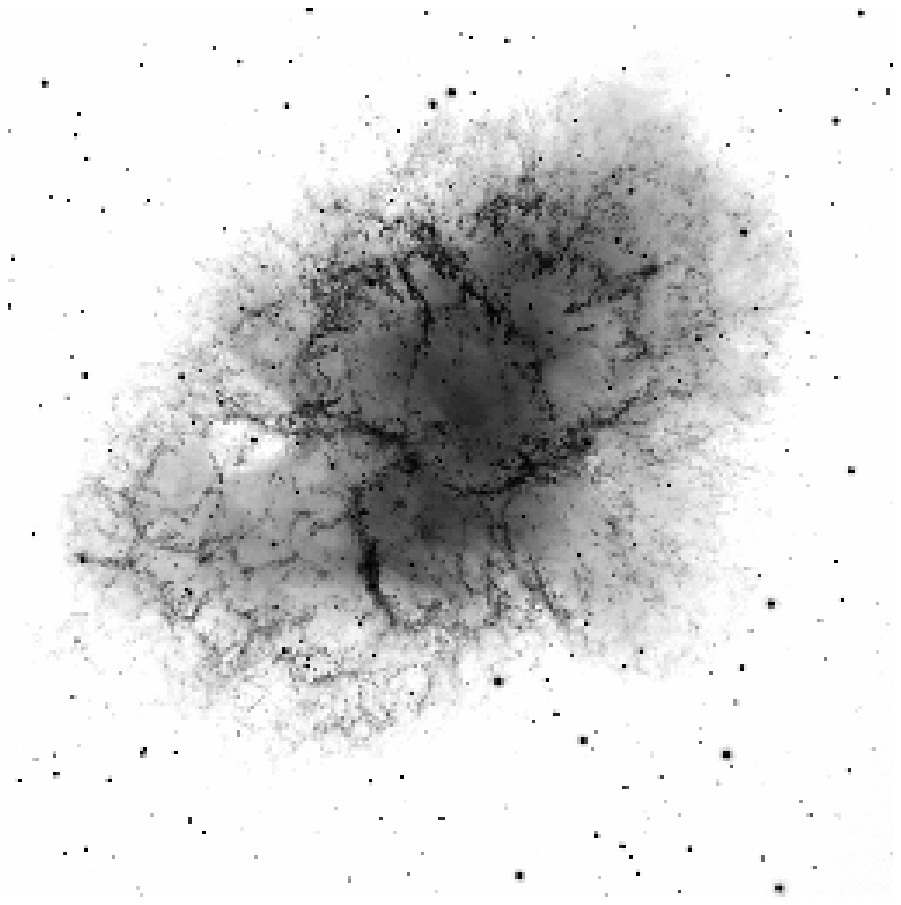}} &
\fbox{\includegraphics[width=0.25\textwidth]{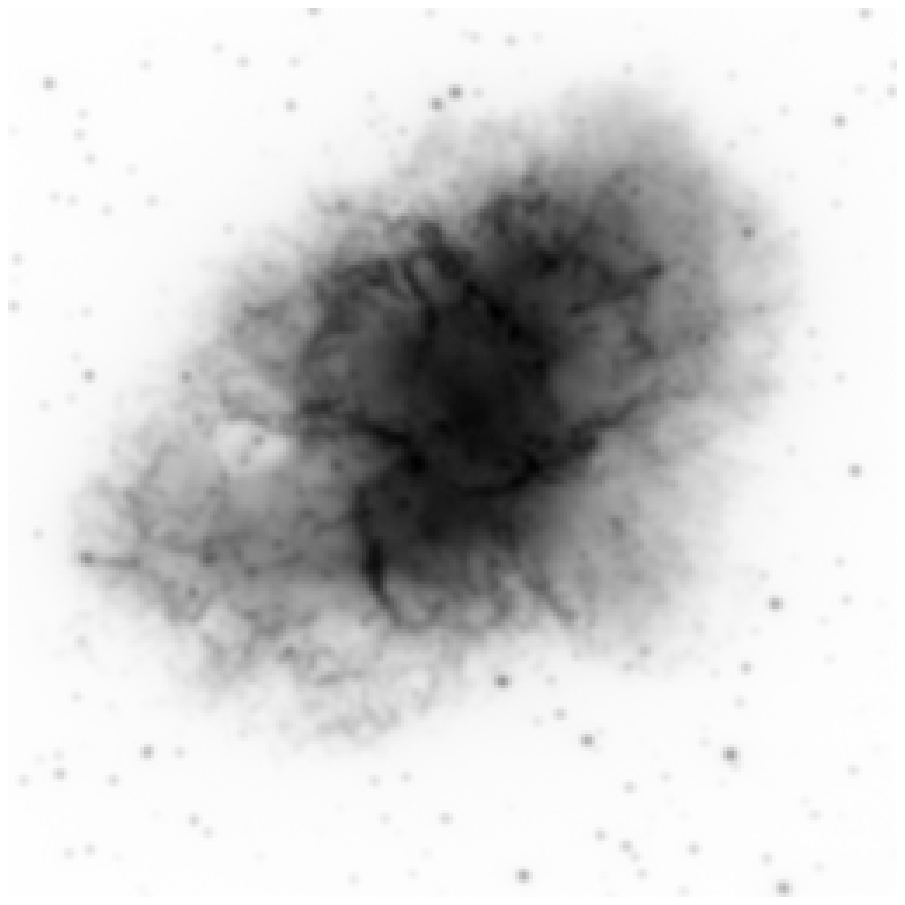}}  &
\fbox{\includegraphics[width=0.25\textwidth]{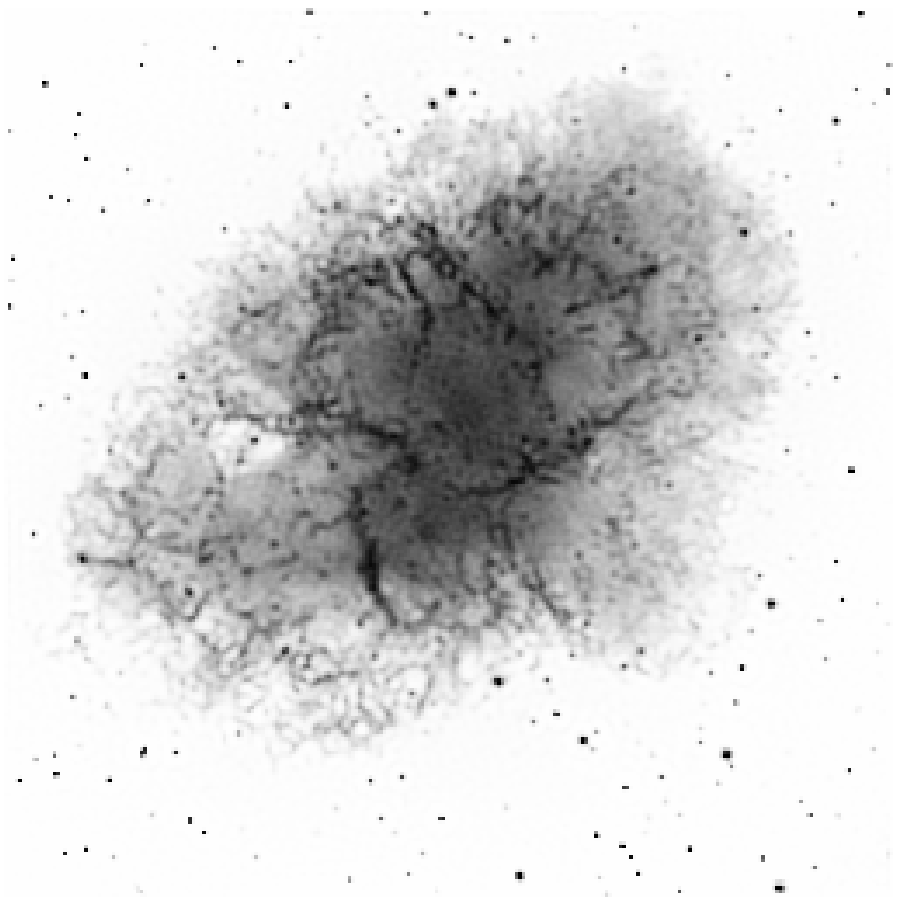}}\\
\fbox{\includegraphics[width=0.25\textwidth]{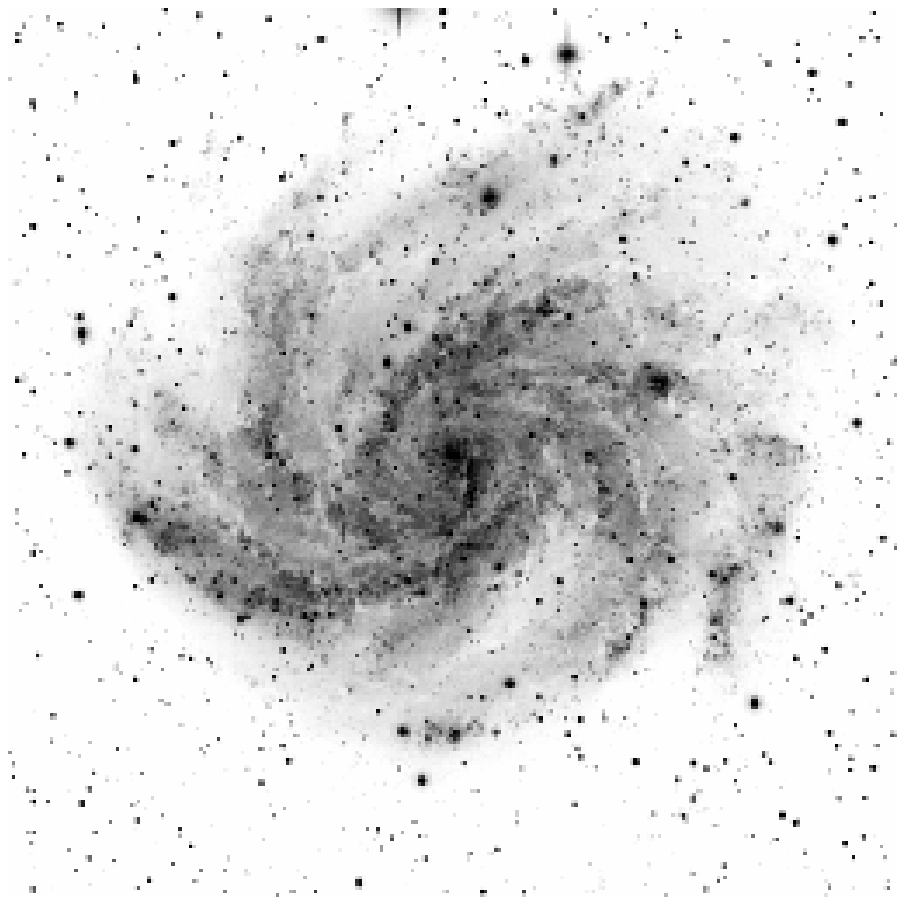}} &
\fbox{\includegraphics[width=0.25\textwidth]{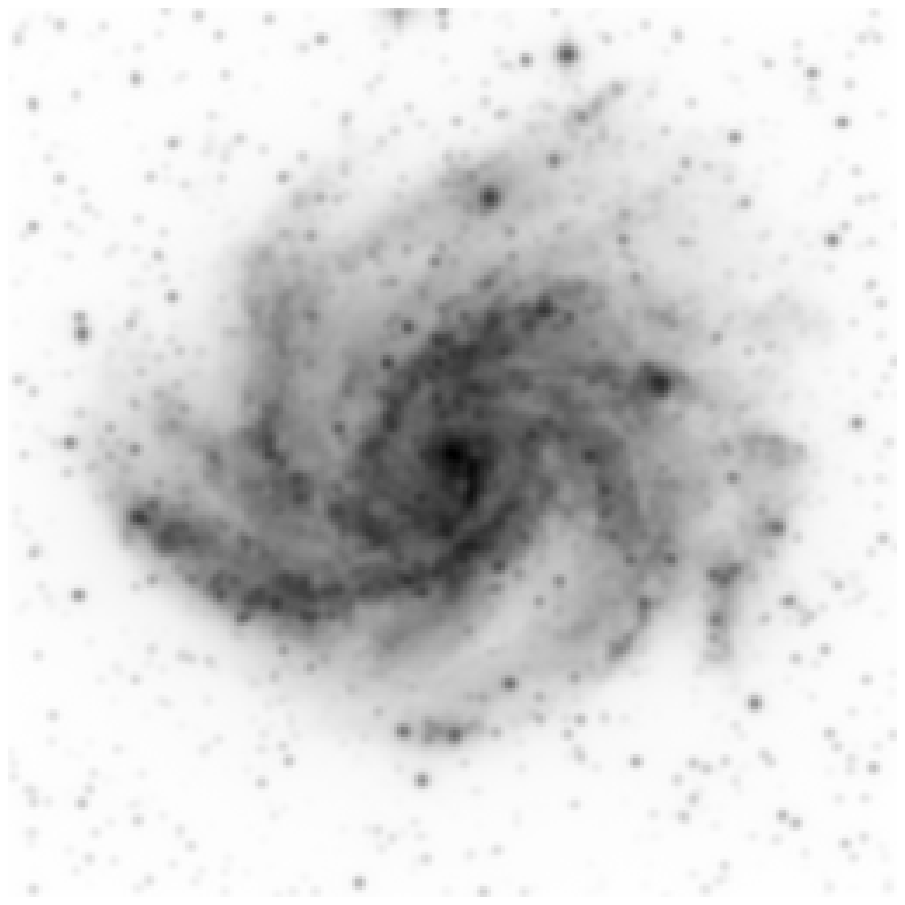}}  &
\fbox{\includegraphics[width=0.25\textwidth]{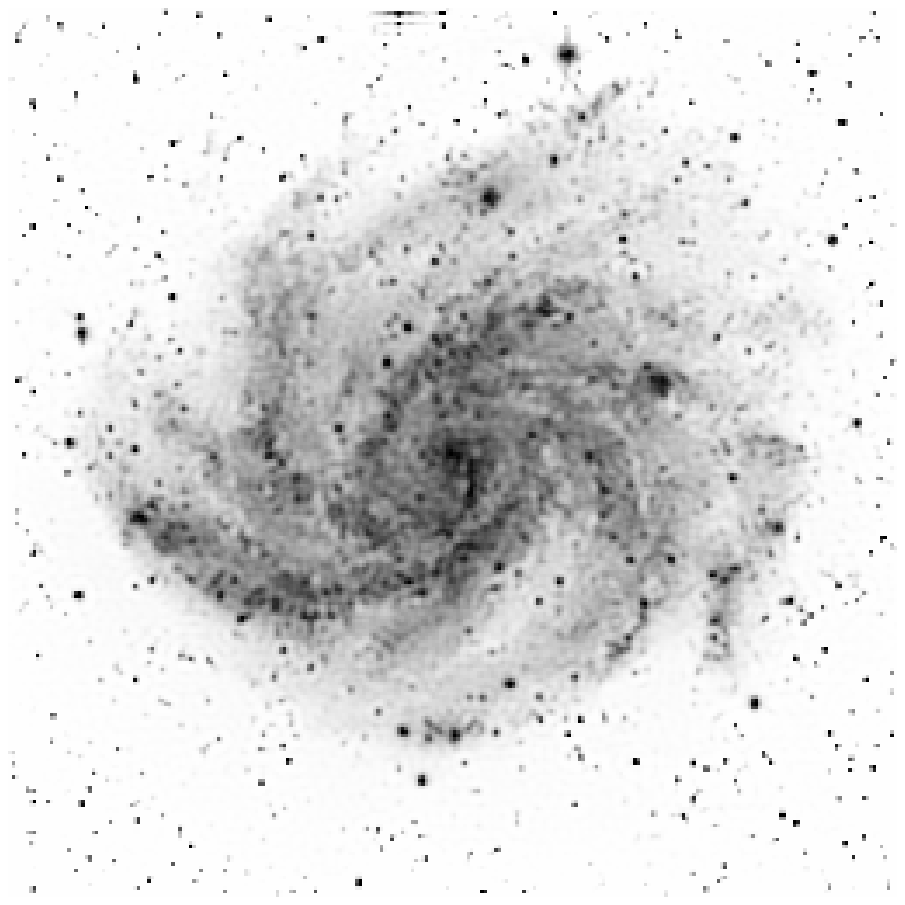}}\\
\fbox{\includegraphics[width=0.25\textwidth]{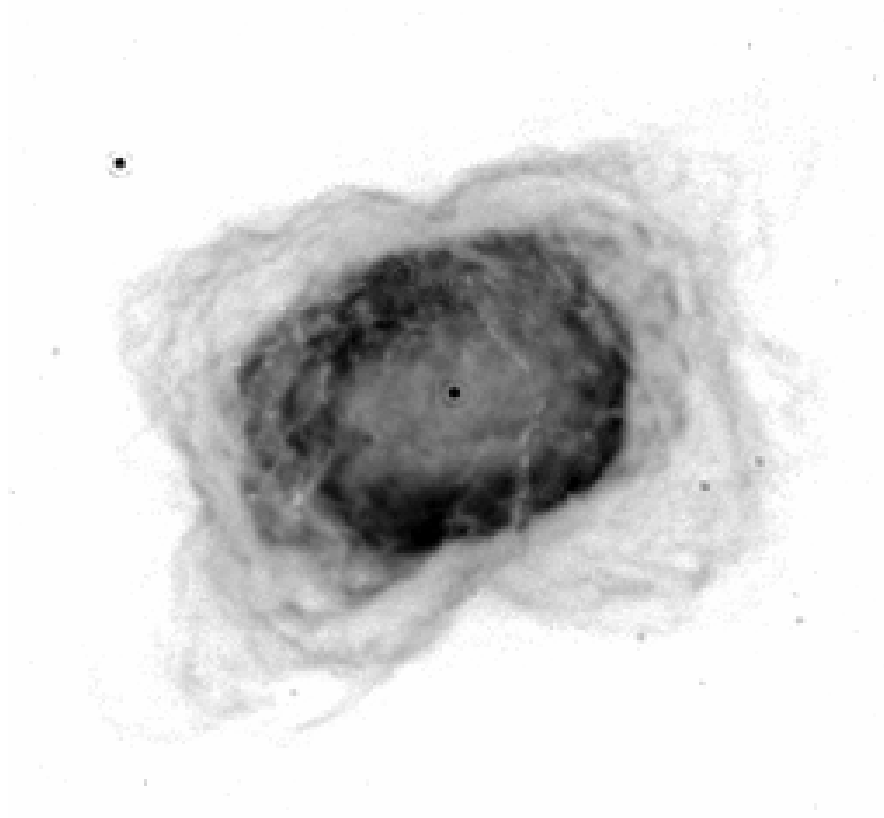}} &
\fbox{\includegraphics[width=0.25\textwidth]{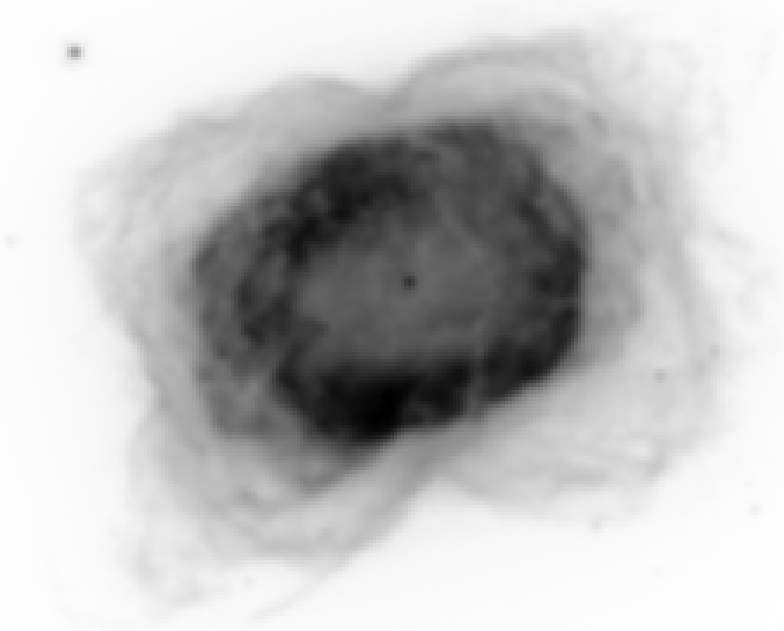}}  &
\fbox{\includegraphics[width=0.25\textwidth]{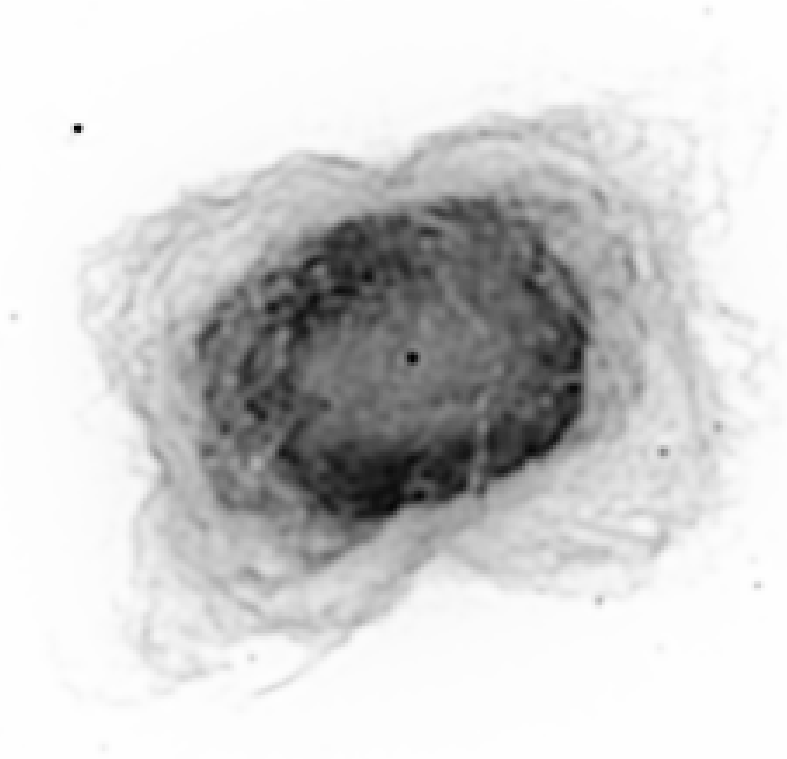}}\\
\end{tabular}
\end{center}
\caption{First column: the objects used for image generation; second column: 
the blurred images in the case of SR=0.67; third column the reconstructed 
objects obtained with our blind algorithm. First row: the Crab nebula NGC 1952, 
second row: the spiral galaxy NGC 6946, third row: the planetary nebula 
NGC 7027.}
\label{fig:diffuse}
\end{figure}

As concerns the initialization of the blind algorithm we use the same PSFs
already used in the case of stellar objects, namely obtained by suitable
autocorrelations of the ideal PSF of the telescope. However in the case
of complex and diffuse objects, as already remarked by other authors (see,
for instance, \cite{biggs}), a difficult and crucial point is the choice 
of the number of inner iterations. We do not have a rule which can be 
successfully applied to all cases as for the stellar objects, i.e. 
$(n_f,n_h)=(50,1)$. By several attempts we find ``best'' numbers for each
case, but it is obvious that this is not a satisfactory situation. For
instance, in the case of the Crab nebula we find $(n_f,n_h)=(13,22)$ for 
SR=0.67 and $(n_f,n_h)=(13,27)$ for SR=0.40, if we search for minimum RMSE on 
the object using 100 outer iterations. Moreover, even if the algorithm is 
convergent, the limit is not, in general, a sensible solution: a suitable 
stopping of the outer iterations is required. In other words the
algorithm is semi-convergent \cite{natterer,bb} as concerns the outer
iterations, i.e. the RMSE on the object first decreases, reaches a minimum
and then goes away. We do not have a proof of this feature which derives
from the numerical experiments. It is obvious that such a situation is not 
satisfactory and we briefly discuss this point in the next section (see also 
the Introduction).

In Table \ref{tab:tab3} we report the best results we have obtained while in 
the third column of Fig. 4 we show the reconstructions of the objects provided
by the blind algorithm. We stress the case of the planetary nebula: it seems that in
this case the algorithm is unable to improve the reconstruction with
respect to that provided by the initial guess. We also remark that the error
on the reconstructed PSF depends on the object: for instance, for the galaxy
it is larger than that for the Crab.

We conclude by reporting an experiment intended to check the effect of an 
underestimated or overestimated SR in the reconstruction of a diffuse object. 
We consider the case of the Crab nebula and PSF with SR=0.67, and we apply our algorithm 
with SR=0.6 and SR=0.8. In the first case, the minimum RMSE on the object and the PSF 
are equal to 13\% and 8.2\%, while in the other the errors on object and PSF are 12\% and 
6.7\%, which are essentially the same values obtained with the correct SR. 
The small variance observed suggests that, in presence of complex objects, the 
availability of a correct SR does not represent a crucial point. In this case, an 
overestimate of the SR value seems to be preferable.

\section{Concluding remarks and perspectives}

In this paper we propose a blind algorithm, based on SGP, for the reconstruction
of astronomical images acquired by a telescope equipped with an AO system.
The algorithm can be classified as an inexact alternating minimization of
the KL divergence depending on both the object and the PSF. The crucial
point is the introduction of different constraints on the object and PSF and
this can be done in a correct way by using the particular structure of SGP.
Moreover the convergence of the algorithm is derived from general results
proved in \cite{bonettini2011} on the convergence of inexact AM.

Since the problem is non-convex, the limit points of the sequence of
iterates may depend on several parameters and, more specifically, on the
initialization of the outer iterations and on the numbers of inner iterations.
In the case of stellar (point-wise) objects we have rules for the
initialization and the numbers of inner iterations which seem to be 
suitable for all
cases we have considered. Obviously, the effectiveness of the approach must
be tested by application to a much broader set of simulated images in a
wide program of simulations (see, for instance, Chap. 12 of \cite{bb})
as well as to real images. We also observe that our blind approach can be
possibly used in conjunction with specific codes, such as the so-called
{\it StarFinder} \cite{diolaiti}, developed for accurate photometric and
astrometric analysis of star clusters. These codes contain a method for 
extracting the PSF from the image of the star field; this PSF can be 
compared with and/or replaced by that provided by our blind approach; also 
in this case the analysis of simulated star fields, in particular crowdy 
star fields, can help to understand when the blind approach is required.

As already remarked in the Introduction, the situation is not so clear in 
the case of more complex astronomical targets. Our preliminary simulations
indicate that the proposed blind method has the semi-convergent property
as concerns the outer iterations (the numbers of the inner iterations are 
fixed by the user and, in any case, they should not be too large).
The difficulties found in optimizing the numbers of inner iterations may 
reside in the fact that in this paper we do not introduce regularization in 
the objective function. Due to the flexibility of SGP the method
can be easily generalized to differentiable regularizers (one only needs
to compute the gradient of the penalty and its positive part) and this 
generalization will be the subject of future work. We stress again that 
the crucial point is to identify regularizers which are suitable for 
specific classes of astronomical targets as well as regularizers which are 
suitable for AO corrected PSFs.

The codes of the algorithms presented and used in this paper are available 
under request.

\section*{Acknowledgments}

This work has been partially supported by MIUR (Italian
Ministry for University and Research), PRIN2008 ``Optimization Methods and
Software for Inverse Problems'', grant 2008T5KA4L, and FIRB - Futuro in Ricerca 2012 
``Learning meets time: a new computational approach for learning in dynamic systems'', 
contract RBFR12M3AC\_002, and by INAF (National
Institute for Astrophysics) under the contract TECNO-INAF 2010 ``Exploiting
the adaptive power: a dedicated free software to optimize and maximize the 
scientific output of images from present and future adaptive optics 
facilities''. The Italian GNCS - INdAM (Gruppo Nazionale per il Calcolo 
Scientifico - Istituto Nazionale di Alta Matematica) is also acknowledged.


\section*{References}
\begin{thebibliography}{100}

\bibitem{barrett} Barrett H H and Meyers K J 2003 {\it Foundations of Image
Science} (New York: Wiley and Sons)

\bibitem{Barzilai-Borwein-1988} Barzilai J and Borwein J M 1988 Two point 
step size gradient methods {\em IMA J. Numer. Anal.} {\bf 8} 141--8

\bibitem{benvenuto2010} Benvenuto F, Zanella R, Zanni L and Bertero M 2010
Nonnegative least-squares image deblurring: improved gradient projection
approaches {\it Inverse Problems} {\bf 26} 025004

\bibitem{bb} Bertero M and Boccacci P 1998 {\it Introduction to Inverse
Problems in Imaging} (Bristol: IoP Publishing)

\bibitem{bertero1998} Bertero M, Bindi D, Boccacci P, Cattaneo M, EVA C
and Lanza 1998 A novel blind deconvolution method with an application
to seismology {\it Inverse Problems} {\bf 14} 815--33

\bibitem{bertero2009} Bertero M, Boccacci P, Desider\`a G and Vicidomini G
2009 Image deblurring with Poisson data: from cells to galaxies 
{\it Inverse Problems} {\bf 25} 123006

\bibitem{bertero2011} Bertero M, Boccacci P, La Camera A, Olivieri C and
Carbillet M 2011 Imaging with LINC-NIRVANA the Fizeau interferometer of
the Large Binocular Telescope; state of the art and open problems
{\it Inverse Problems} {\bf 27} 113001

\bibitem{Bertsekas}
Bertsekas D P 1999 {\em Nonlinear Programming} (Belmont: Athena Scientific)

\bibitem{Bertsekas-Tsitsiklis88}
Bertsekas D P and Tsitsiklis J 1988 {\em Parallel and Distributed Computation: 
Numerical Methods} (Belmont: Prentice-Hall)

\bibitem{biggs} Biggs D S C and Andrews M 1998 Asymmetric iterative
blind deconvolution of multi-frame images {\it Proc. SPIE} {\bf 3461}
328--38

\bibitem{bonettini2009}
Bonettini S, Zanella R and Zanni L 2009 A scaled gradient projection method 
for constrained image deblurring {\em Inverse Problems} {\bf 25} 015002


\bibitem{bonettini2010} Bonettini S and Prato M 2010 Nonnegative image 
reconstruction from sparse Fourier data: a new deconvolution algorithm 
{\em Inverse Problems} {\bf 26} 095001

\bibitem{bonettini2011a} Bonettini s and Ruggiero V 2011 An alternating
extragradient method for total variation-based image reconstruction from
Poisson data {\em Inverse Problems} {\bf 27} 095001 

\bibitem{bonettini2011}
Bonettini S 2011 Inexact block coordinate descent methods with application 
to the nonnegative matrix factorization {\em IMA J. Numer. Anal.} 
{\bf 31} 1431--52

\bibitem{carbillet2004a}
Carbillet M, V\'erinaud C, Femen\'ia B, Riccardi A and Fini L 2004 
Modeling astronomical adaptive optics - I. The software package CAOS
{\em Mon. Not. R. Astron. Soc.} {\bf 356} 1263--75

\bibitem{Chan-Wong-2000}
Chan T F and Wong C K 2000 Convergence of the alternating minimization 
algorithm for blind deconvolution {\em Linear Algebra Appl.} {\bf 316} 259--85

\bibitem{Dai-Fletcher-2005}
Dai Y H and Fletcher R 2006 On the asymptotic behaviour of some new gradient 
methods {\em Math. Program.} {\bf 103} 541--59

\bibitem{Dai-Fletcher-2006}
Dai Y H and Fletcher R 2006 New algorithms for singly linearly constrained 
quadratic programming problems subject to lower and upper bounds 
{\em Math. Program.} {\bf 106} 403--21

\bibitem{Dai-etal-2006}
Dai Y H, Hager W W, Schittkowski K and Zhang H 2006 The cyclic 
Barzilai--Borwein method for unconstrained optimization 
{\em IMA J. Numer. Anal.} {\bf 26} 604--27

\bibitem{desidera2006}
Desider\`{a} G, Anconelli B, Bertero M, Boccacci P and Carbillet M 2006 
Application of iterative blind deconvolution to the reconstruction of LBT 
LINC--NIRVANA images {\em Astron. Astrophys.} {\bf 452} 727-–34

\bibitem{desidera2009}
Desider\`{a} G and Carbillet M 2009 Strehl-constrained iterative blind 
deconvolution for post-adaptive-optics data {\em Astron. Astrophys.} 
{\bf 507} 1759–-62

\bibitem{diolaiti} Diolaiti E, Bendinelli O, Bonaccini D, Close L M, Currie 
D G and Parmeggiani G 2000 StarFinder: an IDL GUI based code to analyze
crowded fields with isoplanatic correcting PSF fitting {\em Proc. SPIE}
{\bf 4007} 879--88

\bibitem{fish} Fish D A, Brinicombe A M and Pike E R 1995 Blind 
deconvolution by means of the Richardson-Lucy algorithm 
{\it J. Opt. Soc. Am.} {\bf A-12} 58--65

\bibitem{Frassoldati-etal-2008}
Frassoldati G, Zanghirati G and Zanni L 2008 New adaptive stepsize 
selections in gradient methods {\em J. Ind. Manag. Optim.} {\bf 4} 299--312

\bibitem{Grippo-Sciandrone-1999}
Grippo L and Sciandrone M 1999 Globally convergent block-coordinate 
techniques for unconstrained optimization {\em Optim. Method Softw.} 
{\bf 10} 587--637

\bibitem{Grippo-Sciandrone-2000}
Grippo L and Sciandrone M 2000 On the convergence of the block nonlinear 
Gauss--Seidel method under convex constraints {\em Oper. Res. Lett.} 
{\bf 26} 127--36

\bibitem{Hansen-etal-2006}
Hansen P C, Nagy J C and O'Leary D P 2006 {\em Deblurring Images. Matrices, 
Spectra, and Filtering} (Philadelphia: SIAM)

\bibitem{holmes} Holmes T J 1992 Blind deconvolution of quantum-limited
incoherent imagery: maximum-likelihood approach {\em J. Opt. Soc. Am.}
{\bf A-9} 1052--61

\bibitem{jefferies}
Jefferies S M and Christou J C 1993 Restoration of astronomical images by 
iterative blind deconvolution {\em Astrophys. J.} {\bf 415} 862--74

\bibitem{lanteri1994} Lant\'eri H, Aime C, Beaumont H and Gaucherel P 1994
Blind deconvolution using the Richardson-Lucy algorithm {\em Proc. SPIE}
{\bf 2312} 182--92 

\bibitem{lee2001} Lee D D and Seung H S 2001 Algorithms for non-negative
matrix factorization {\it Advances in Neural Information Processing 13
(Proc. NIPS$^*$2000)} (MIT Press) 556--62

\bibitem{levin}
Levin A Weiss Y Durand F and Freeman W T 2009 Understanding and evaluating 
blind deconvolution algorithms {\em IEEE Conf. Computer Vision and Pattern 
Recognition} 1964--71

\bibitem{LuoTseng91}
Luo Z-Q and Tseng P 1992 On the convergence of the coordinate descent method 
for convex differentiable minimization {\em J. Optimiz. Theory App.} 
{\bf 72} 7--35

\bibitem{natterer} Natterer F and W\"ubbeling F 2001 {\it Mathematical Methods
in Image Reconstruction} (Philadelphia: SIAM)

\bibitem{Nocedal} Nocedal J and Wright S J 2006 {\it Numerical Optimization:
Second Edition} (New York: Springer)

\bibitem{Powell}
Powell M J D 1973 On search directions for minimization algorithms 
{\em Math. Program.} {\bf 4} 193--201

\bibitem{Prato-etal-2012}
Prato M, Cavicchioli R, Zanni L, Boccacci P and Bertero M 2012 Efficient 
deconvolution methods for astronomical imaging: algorithms and IDL--GPU 
codes {\em Astron. Astrophys.} {\bf 539} A133

\bibitem{snyder1993} Snyder D L, Hammoud A M and White R L 1993 Image recovery 
from data acquired with a charge-coupled-device camera {\em J. Opt. Soc. Am.}
{\bf A-10} 1014--23

\bibitem{snyder1995}
Snyder D L, Helstrom C W, Lanterman A D, Faisal M and White R L 1995 
Compensation for readout noise in ccd images 
{\em J. Opt. Soc. Am.} {\bf A-12} 272--83

\bibitem{stagliano} Staglian\`o, Boccacci P and Bertero M 2011 Analysis of
an approximate model for Poisson data reconstruction and a related
discrepancy principle {\em Inverse Problems} {\bf 27} 125003

\bibitem{Tseng91}
Tseng P 1991 Decomposition algorithm for convex differentiable minimization 
{\em J. Optimiz. Theory App.} {\bf 70} 109--35

\bibitem{tsumuraya} Tsumuraya F, Miura N and Baba N 1994 Iterative blind
deconvolution method using Lucy's algorithm {\it Astron. Astrophys.}
{\bf 282} 699--708

\bibitem{zanella} Zanella R, Boccacci P, Zanni L and Bertero M 2009 Efficient 
gradient projection methods for edge-preserving removal of Poisson noise
{\em Inverse Problems} {\bf 25} 045010

\bibitem{Zhou-etal-2006}
Zhou B, Gao L and Dai Y H 2006 Gradient methods with adaptive step-sizes 
{\em Comput. Optim. Appl.} {\bf 35} 69--86

\end{thebibliography}
\end{document}